\documentclass[journal]{IEEEtran}
\usepackage{times, bm}
\usepackage[english]{babel}
\usepackage{amsfonts}
\usepackage{graphics}
\usepackage{mathrsfs}
\usepackage{eufrak}
\usepackage{amsbsy}
\usepackage{amssymb}
\usepackage{latexsym}
\usepackage[T1]{fontenc}
\usepackage{amsmath}
\usepackage[dvips]{graphicx}
\usepackage{verbatim, cite}
\usepackage[normalem]{ulem}
\usepackage{indentfirst}
\usepackage{cite}
\usepackage{subfigure}
\usepackage{amssymb}
\usepackage{amsmath}
\usepackage{multicol}
\usepackage{amsfonts}
\usepackage{geometry}
\usepackage{times}
\usepackage[dvips]{graphicx}
\usepackage{fancybox}
\usepackage{url}
\usepackage{bm}
\usepackage{dsfont}
\usepackage{stfloats}
\usepackage{enumerate}

\newcommand{\figw}{0.9\linewidth}
\newcommand{\ben}{\begin{enumerate}}
\newcommand{\een}{\end{enumerate}}
\newcommand{\be}{\begin{equation}}
\newcommand{\ee}{\end{equation}}
\newcommand{\bea}{\begin{eqnarray}}
\newcommand{\eea}{\end{eqnarray}}
\newcommand{\bc}{\begin{cases}}
\newcommand{\ec}{\end{cases}}
\newcommand{\bi}{\begin{itemize}}
\newcommand{\ei}{\end{itemize}}
\newcommand{\e}{\item}
\newcommand{\eq}[1]{(\ref{#1})}

\def\Cap{{\mathrm{C}}}
\def\entropy{{\mathrm{g}}}
\def\dual{{\mathrm{d}}}
\def\Pcn{{P^{c}_{n}}}
\def\Pcd{{P^{c}_{d}}}
\def\fn{{\mathrm{f}_{n}}}
\def\f0{{\mathrm{f}_{0}}}
\def\Gfunc{{\mathrm{G}}}

\def\limsup{{\underset{T \to \infty}{\lim\!\sup\ }}}
\def\liminf{{\underset{T \to \infty}{\lim\!\inf}\ }}

\def\fnpidef{{\limsup \frac{1}{T} \sum_{t=0}^{T-1} \Exp [ \fn(D_{n}(t))]}}

\def\S{{\textbf{S}}}
\def\s{{\textbf{s}}}
\def\o{{\textbf{o}}}
\def\O{{\textbf{O}}}
\def\P{{\mathbf{P}}}
\def\H{{\mathbf{H}}}
\def\Htildebf{{\widetilde{\mathbf{H}}}}
\def\Htilde{{\widetilde{H}}}
\def\h{{\mathbf{h}}}

\def\R{{\mathbf R}}
\def\Exp{{\mathbb E}}

\def\N{{\mathcal N}}
\def\G{{\mathcal G}}
\def\L{{\mathcal L}}
\def\Lag{{\mathfrak L}}
\def\X{{\mathcal X}}

\def\Sset{{\mathcal S}}
\def\Oset{{\mathcal O}}
\def\Hset{{\mathcal H}}

\def\U{{\mathbf U}}
\def\E{{\mathbf E}}

\def\Z{{\mathbf Z}}

\def\D{{\mathbf D}}
\def\r{{\mathbf r}}
\def\d{{\mathbf d}}
\def\p{{\mathbf p}}

\def\muin{{\mu_{*,n}}}
\def\muout{{\mu_{n,*}}}

\def\Pout{{P_{n}}}

\def\Dmax{{D_{\rm max}}}
\def\Dmin{{D_{\rm min}}}

\def\mumax{{\mu_{\rm max}}}
\def\Rmax{{R_{\rm max}}}
\def\RmaxSq{{R^{2}_{\rm max}}}

\def\lmax{{l_{\rm max}}}
\def\Hmax{{H_{\rm max}}}
\def\HmaxSq{{H^{2}_{\rm max}}}
\def\Pmax{{P_{\rm max}}}
\def\PmaxSq{{P^{2}_{\rm max}}}

\def\Pcmax{{\alpha_{n}\Rmax}}
\def\PcmaxSq{{\alpha_{n}^{2}\RmaxSq}}

\def\thetabf{{\boldsymbol \theta}} 

\IEEEoverridecommandlockouts

\usepackage[linesnumbered,lined, algoruled]{algorithm2e}

\newtheorem{thm}{Theorem}[section]

\newtheorem{lem}[thm]{Lemma}
\newtheorem{rem}[thm]{Remark}

\newcommand{\qed}{\nobreak \ifvmode \relax \else
      \ifdim\lastskip<1.5em \hskip-\lastskip
      \hskip1.5em plus0em minus0.5em \fi \nobreak
      \vrule height0.75em width0.5em depth0.25em\fi}

\title{Dynamic Compression-Transmission for Energy-Harvesting Multihop Networks with Correlated Sources}

\author{Cristiano Tapparello\IEEEauthorrefmark{1}, Osvaldo Simeone\IEEEauthorrefmark{2} and Michele Rossi\IEEEauthorrefmark{1}
\thanks{\IEEEauthorrefmark{1}Department of Information Engineering, University of Padova, Italy. \IEEEauthorrefmark{2}CWCSPR, New Jersey Institute of Technology, New Jersey, USA. This work has been supported in part by the FP7 EU project ``SWAP'' G.A.\ no.\ 251557. The work of O. Simeone was partially supported by US NSF under grant CCF-0914899.}}
\begin{document}
\sloppy

\maketitle


\begin{abstract}
Energy-harvesting wireless sensor networking is an emerging technology with applications to various fields such as environmental and structural health monitoring. A distinguishing feature of wireless sensors is the need to perform both source coding tasks, such as measurement and compression, and transmission tasks. It is known that the overall energy consumption for source coding is generally comparable to that of transmission, and that a joint design of the two classes of tasks can lead to relevant performance gains. Moreover, the efficiency of source coding in a sensor network can be potentially improved via distributed techniques by leveraging the fact that signals measured by different nodes are correlated. 

In this paper, a data gathering protocol for multihop wireless sensor networks with energy harvesting capabilities is studied whereby the sources measured by the sensors are correlated. Both the energy consumptions of source coding and transmission are modeled, and distributed source coding is assumed. The problem of dynamically and jointly optimizing the source coding and transmission strategies is formulated for time-varying channels and sources. The problem consists in the minimization of a cost function of the distortions in the source reconstructions at the sink under queue stability constraints. By adopting perturbation-based Lyapunov techniques, a close-to-optimal online scheme is proposed that has an explicit and controllable trade-off between optimality gap and queue sizes. The role of side information available at the sink is also discussed under the assumption that acquiring the side information entails an energy cost. It is shown that the presence of side information can improve the network performance both in terms of overall network cost function and queue sizes. 
\end{abstract}


\section{Introduction} \label{sec:intro}
Wireless sensor networks have found applications in a large number of fields such as environmental sensing and structural health monitoring~\cite{Dargie10}. In such applications, the maintenance necessary to replace the batteries when depleted is often of prohibitive complexity, if not impossible. Therefore, sensors that harvest energy from the environment, e.g., in the form of solar, thermal, vibrational or radio energy~\cite{Paradiso05}~\cite{Conrad08}, have been proposed and are now commercially available. 

Given the interest outlined above, the problem of designing optimal transmission protocols for energy harvesting wireless sensor networks has recently received considerable attention. In the available body of work reviewed below in Section~\ref{sec:prior_work}, the only source of energy expenditure is the energy used for transmission. This includes, e.g., the energy used by the power amplifiers. However, a distinguishing feature of sensor networks is that the sensors have not only to carry out transmission tasks, but also {\it sensing and source coding tasks}, such as compression. The source coding tasks entail a non-negligible energy consumption. In fact, reference~\cite{Barr06} demonstrates that the overall cost required for compression\footnote{This reference considers transmission of Web data.} is comparable with that needed for transmission, and that a joint design of the two tasks can lead to significant energy saving gains. Another distinguishing feature of sensor networks is that the efficiency of source coding can be improved via {\it distributed source coding techniques} (see, e.g.,~\cite{ElGamal12}) by leveraging the fact that sources measured by different sensors are generally correlated (see, e.g.,~\cite{Zordan11}).

\subsection{Contributions}
\label{sec:contributions}
In this paper, we focus on an energy-harvesting wireless sensor network and account for the energy costs of both source coding and transmission. Moreover, we assume that the sensors can perform distributed source coding to leverage the correlation of the sources measured at different sensors. A key motivation for enabling distributed source coding in energy-harvesting networks is that this enables sensors with correlated measurements to trade energy resources among them, to an extent determined by the amount of correlation. For instance, a sensor that is running low on energy can benefit from the energy potentially available at a nearby node if the latter has correlated measurements. This is because, through distributed source coding, the transmission requirements on the first sensor are eased by the transmission of correlated information from the nearby sensor.

We study the problem of dynamically and jointly optimizing the source coding and transmission strategies over time-varying channels and sources. The problem consists in the minimization of a cost function of the distortions in the source reconstructions at the sink under queue stability constraints. Our approach is based on the Lyapunov optimization strategy with weight perturbation developed in~\cite{Huang10}. We devise an efficient online algorithm that only takes actions based on the harvested energy, on the current state of channel, queues and energy reserves, and also based on the statistical description of the source correlation. We prove that the proposed policy achieves an average network cost that can be made arbitrarily close to the optimal one with a controllable trade-off between the sizes of the queues and batteries. 

We also investigate the role of side information available at the sink under the assumption that acquiring the side information entails an energy cost. It is shown that properly allocating the available (harvested) energy to both the tasks of transmission and side information measurement has significant benefits both in terms of overall network cost function and queue sizes.

\subsection{Prior Work}
\label{sec:prior_work}

We start by introducing related prior work that assumes energy harvesting. The literature on this topic is quickly increasing in volume but it mostly (with the exception of~\cite{Castiglione11}) accounts only for the energy consumption of the transmission component, and does not model the contribution of the source coding part. In this context, references~\cite{Ozel11} and~\cite{Keong11} studied the problem of maximizing the throughput or minimizing the completion time for a single link energy-harvesting system by focusing on both offline and online policies (see also~\cite{Devillers11,Chen11}). A related work is also reference~\cite{Sharma10} that finds a power allocation policy that stabilizes the data queue whenever feasible. Still, for a point-to-point system, using large deviation tools, the effect of finite data queue length and battery size is studied in~\cite{Srivastava10} in terms of scaling results as the battery and queue grow large. 
We now consider work on multihop energy-harvesting networks. As mentioned above, all the works at hand only account for the energy used for transmission. Moreover, source correlations and distributed source coding are not accounted for. In~\cite{Huang10} assuming independent and identically distributed (i.i.d.) channel states and energy harvesting processes, a Lyapunov optimization technique with weight perturbation~\cite{Huang11} is leveraged to obtain approximately optimal strategies in terms of a general function of the data rates under queue stability constraints. The proposed technique obtains an explicit trade-off in terms of data queue length and battery size. An extension of this work that assumes more general arrival, channel state and recharge processes along with finite batteries and queues is put forth in~\cite{Mao11}. Also related are~\cite{Gratzianas10},~\cite{Lin07} and~\cite{Lin07_2} that tackle similar problems.



We now discuss work that accounts for the energy trade-offs related to source coding and transmission. These works (except~\cite{Castiglione11}) do not model the additional constraints arising from energy harvesting. Moreover, they do not allow for distributed source coding. The joint design of source coding and transmission parameters is investigated through various algorithms, for either static scenarios in~\cite{Luna03,Akyol08} or dynamic scenarios in~\cite{Neely08,Sharma09}. Specifically, references~\cite{Neely08} and~\cite{Sharma09} studied the trade-offs between energy used for compression, or more generally source coding, and transmission by assuming i.i.d. source and channel processes and arbitrarily large data buffer. Using Lyapunov optimization techniques, a policy with close-to-optimal power expenditure and an explicit trade-off with the delay is derived for a given average distortion. The problem of optimal energy allocation between source coding and transmission for a point-to-point system was studied in~\cite{Castiglione11}.


Finally, distributed source coding techniques for multihop sensor networks has been studied in~\cite{Cristescu06} and~\cite{Cui07}. In~\cite{Cristescu06}, the problem of optimizing the transmission and compression strategy was tackled under distortion constraints in a centralized fashion, whereas~\cite{Cui07} proposes a distributed algorithm that maximizes an aggregate utility measure defined in terms of the distortion levels of the sources. Both these works do not consider energy harvesting nor the energy consumption of the sensing process.

\subsection{Paper organization}
The rest of the paper is organized as follows. In Section~\ref{sec:system_model} we present the system model and we state the optimization problem. In Section~\ref{sec:optimal_penalty} we obtain a lower bound on the optimal network cost for the proposed problem. In Section~\ref{sec:proposed_algorithm} we present the proposed  algorithm designed following the Lyapunov optimization framework and we show how it can be implemented in a distributed fashion. Section~\ref{sec:performance} formalizes the main results of our paper and provide analytical insights into the performance of the proposed policy. Section~\ref{sec:side_information} proposes an extended version of the problem, where the sink node acts as a cluster head that is able to acquire correlated side information to improve the system performance. In Section~\ref{sec:results} we prove the effectiveness of our analytical analysis and discuss the impact of the optimization parameters. Section~\ref{sec:conclusions} concludes the paper.

\section{System model}
\label{sec:system_model}
\begin{figure}[t!]
\begin{center}
\includegraphics[width=\figw]{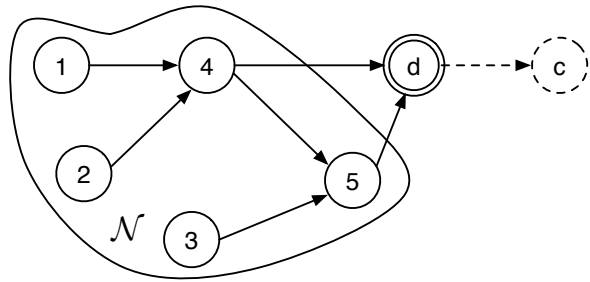}
\caption{A set $\N$ of energy-harvesting nodes communicate correlated sources to a destination $d$. For the more general model of Section~\ref{sec:side_information}, the destination $d$ acts as a cluster head and communicates to a network collector node (shown in dashed lines). In this latter model, the node $d$ can collect side information correlated to the sources measured by the nodes.}
\label{fig:scenario}
\end{center}
\end{figure}
We consider a wireless network modeled by a direct graph $\G = (\N \cup \{ d \}, \L)$, where $\N = \left \{ 1, 2, \ldots, N\right \}$ is the set of nodes in the network, $d$ is the destination (or sink), and $\L \subset \{ (n,m)\text{: } n,m \in \N \cup \{d\}, n \neq m \}$ represents the set of communication links, see Fig.~\ref{fig:scenario} for an illustration. We define $\lmax$ as the maximum number of transmission links that any node $n \in \N \cup \{d\}$ can have. As discussed below, we allow for fairly general interference models. We will consider a more general model in Section~\ref{sec:side_information} in which the sink acts as a cluster head for the set of nodes $\N$, and reports to a collector node $c$ (see Fig.~\ref{fig:scenario}). 

\subsection{Transmission Model}
\label{sec:transmission_model}
The transmission model follows the framework of, e.g.,~\cite{Georgiadis06}. According to this model, the network operates in slotted time and, at every time slot $t = 1,2, \ldots$, each node $n \in \N$ allocates power $P_{n,m}(t) \geq 0$ to each outgoing link $(n,m) \in \L$ for data transmission. In what follows, we refer to the number of {\it channel uses} (or transmission symbols) per time slot as the baud rate multiplied by the slot duration. At the generic time slot $t$ we define $\P(t)=\{P_{n,m}(t)\}$, with $(n,m) \in \L$, as the power allocation matrix and the total transmission power of node $n$, that is
\be
P_{n}(t) = \sum_{m\text{: } (n,m) \in \L} P_{n,m}(t),
\label{eq:energy_out}
\ee
which is assumed to satisfy the constraint $P_{n}(t) \leq \Pmax$, for some $\Pmax < \infty$.
The transmission rate $\mu_{n,m}(t)$ on link $(n,m)$ depends on the power allocation matrix $\P(t)$ and on the current {\it channel state} $\S(t)=\{S_{n,m}(t)\}$ with $(n,m) \in \L$. The latter accounts, for instance, for the current fading channels or for the connectivity conditions on the network links. We assume that $\S(t)$ takes values in some finite set $\Sset = (\s_{1},\s_{2},\ldots,\s_{M})$, is constant within a time slot, but is independent and identically distributed (i.i.d.) across time slots. We use $\rho_{\s_{i}}=\Pr \left [ \S(t) = \s_{i} \right]$ for $i = 1, \ldots, M$. We write
\be
\mu_{n,m}(t) = \Cap_{n,m}(\P(t),\S(t)),
\label{eq:power_rate}
\ee
where $\Cap_{n,m}(\P(t),\S(t))$ is the capacity-power curve for link $(n,m)$ expressed in terms of bits per channel use (transmission symbol). The latter depends on the specific network transmission strategy, which includes the modulation and coding/decoding schemes used on all links. We assume that function $\Cap_{n,m}(\P(t),\S(t))$ is continuous in $\P(t)$ and non decreasing in $P_{n,m}(t)$. An example of the function $\Cap_{n,m}(\P(t),\S(t))$ is the Shannon capacity obtained by treating interference as noise at the receivers, namely
\be
\label{eq:capacity_power}
\begin{split} \Cap_{n,m}(\P&(t),\S(t)) \propto \log \left (1 + \frac{P_{n,m}(t)S_{n,m}(t)}{N_{0} + \sum_{(l,n) \in \L}P_{l}(t)S_{l,n}(t)} \right ),
\end{split}
\ee
where $S_{n,m}(t)$ represents the channel power gain on link $(n,m)$ and $N_{0}$ is the noise spectral density. We assume that there exists some finite constant $\mumax$ such that $\mu_{n,m}(t) \leq \mumax$ for all $t$, any power allocation vector $\P(t)$ and  channel state $\S(t)$. Moreover, following~\cite{Huang10}, we assume that the function $\Cap_{n,m}(\P(t),\S(t))$ satisfies the following properties:

{\it Property 1:} For any power allocation matrix $\P(t)$, we have:
\be
\Cap_{n,m}(\P(t),\S(t)) \leq \xi P_{n,m}(t),
\ee
for some finite constant $\xi >0$;

{\it Property 2:} For any power allocation matrix $\P(t)$ and matrix $\P^{\prime}(t)$ obtained by $\P(t)$ by setting the entry $P_{n,m}(t)$ to zero for a given $(n,m)$ pair, we have:
\be
\Cap_{a,b}(\P(t),\S(t)) \leq \Cap_{a,b}(\P^{\prime}(t),\S(t)), 
\ee
for all $(a,b) \in \L$, with $(a,b) \neq (n,m)$.

Note that both properties are satisfied by typical choices of function $\Cap_{n,m}(\P(t),\S(t))$ such as~\eq{eq:capacity_power}. In fact, {\it Property 1} is satisfied if function $\Cap_{n,m}(\P(t),\S(t))$ is concave with respect to $P_{n,m}(t)$, while {\it Property 2} states that interference due to power spent on other links cannot be beneficial.\footnote{This may not be the case if sophisticated physical layer techniques are used, such as successive interference cancelation (see, e.g.,~\cite{ElGamal12}).} Finally, we define the total outgoing transmission rate $\muout(t)$ from a node $n \in \N$ at time $t$ as 
\be
\muout(t) = \sum_{m\text{: }(n,m) \in \L} \mu_{n,m}(t),
\label{eq:rate_out}
\ee
and the total incoming transmission rate $\muin(t)$ at a node $n \in \N$ as
\be
\muin(t) = \sum_{m\text{: }(m,n) \in \L} \mu_{m,n}(t).
\label{eq:rate_in}
\ee

\subsection{Data Acquisition, Compression and Distortion Model}
\label{sec:distortion_model}
At each time slot, each node of the network is able to sense the environment and to acquire spatially correlated measurements. The measurements are then routed through the network to be gathered by a sink node, as illustrated in Fig.~\ref{fig:scenario}. Before transmission, the acquired data is compressed via adaptive lossy source coding by leveraging the spatial correlation of the measurements. Specifically, we define the {\it source state} at time $t$ as the spatial correlation matrix describing the signal within this time slot, which is referred to as  $\O(t)=\{O_{n,m}(t)\}$ with $n,m \in \N$. We assume that $\O(t)$ takes values in some finite set $\Oset = \{ \o_{1}, \o_{2}, \ldots, \o_{L}\}$, remains constant within a time slot, but is i.i.d. across time slots. Additionally, we define the mdf  $\rho_{\o_{i}} = \Pr[\O(t) = \o_{i}]$.
Each node $n \in \N$ compresses the measured source with rate $R_{n}(t) \leq \Rmax$ bits per source symbol and targets a reproduction distortion at the sink of $\Dmin \leq D_{n}(t) \leq \Dmax$, with $0 < \Rmax, \Dmin \leq \Dmax < \infty$. Note that imposing a strictly positive lower bound on $D_{n}(t)$ is without loss of generality because the rate $R_{n}(t)$ is upper bounded by a finite constant and therefore the distortion $D_{n}(t)$ cannot in general be made arbitrarily small (see, e.g.,~\cite{ElGamal12}). The distortion is measured according to some fidelity criterion such as mean square error (MSE). We define the rate vector as $\R(t) = (R_{1}(t),\ldots,R_{N}(t))$ and the distortion vector as $\D(t) = (D_{1}(t), \ldots, D_{N}(t))$. Due to the spatial correlation of the measurements, {\it distributed source coding techniques} can be leveraged. Thanks to these techniques, the rates of different users can be traded without affecting the achievable distortions, to an extent that depends on the amount of spatial correlation~\cite{ElGamal12}. The adoption of distributed source coding entails that, given certain distortion levels $\D(t)$, the rates $\R(t)$ can be selected arbitrarily as long as they satisfy appropriate joint constraints. Under such constraints, a sink receiving data at the specified rates is able to recover all sources at the given distortion levels.
\begin{figure}[t!]
\begin{center}
\includegraphics[width=\figw]{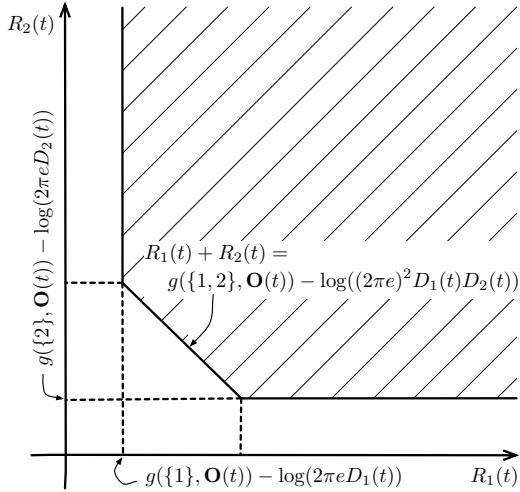}
\caption{Illustration of the rate region~\eq{eq:rates_dist} for correlated sources and $\N=\{1,2\}$. For all rate pairs $(R_{1}(t),R_{2}(t))$, there exists a coding schemes that enables the sink to recover the two sources with distributed distortion (MSE) levels $D_{1}(t)$ and $D_{2}(t)$, respectively.}
\label{fig:rate_region}
\end{center}
\end{figure}

To elaborate on this point, consider the following conditions on the rates $R_{n}(t)$ and distortions $D_{n}(t)$ for $n \in \N$:
\be
\sum_{n \in \X} R_{n}(t) \geq \entropy(\X,\O(t)) - \log \left ( (2 \pi e)^{\vert \X \vert} \prod_{n \in \X} D_{n}(t) \right ),
\label{eq:rates_dist}
\ee
for all $\X \subseteq \N$, where $\entropy(\X,\O(t))$ denotes the joint conditional differential entropy of the sources measured by the nodes in the subset $\X$, where conditioning is with respect to the sources measured by the nodes in the complement $\N \setminus \X$. For instance, for jointly Gaussian sources with zero mean and correlation matrix $\O(t)$, we have
\be
\label{eq:entropy_Gauss}
\entropy(\X,\O(t)) = \frac{1}{2} \log \left (  \frac{\det \O(t) }{\det \O(t)|_{\N \setminus \X} } \right ),
\ee
where $\O(t)|_{\N \setminus \X}$ represents the correlation submatrix of the sources measured by nodes in $\N \setminus \X$. If the rates satisfy conditions~\eq{eq:rates_dist}, it is known~\cite{Zamir99} that, for sufficiently small distortions and any well-behaved joint source distribution, the sink is able to recover all the sources within MSE levels $D_{n}(t)$ for all $n \in \N$. We remark that this conclusion is also valid for any distortion tuple $\D(t)$ if the sources are jointly Gaussian. 

As an example, the rate region for $\N = \{1,2\}$ is sketched in Fig.~\ref{fig:rate_region}. The rates $R_{1}(t)$ and $R_{2}(t)$ at which the two source sequences are acquired and compressed at the two nodes can be traded with one another without affecting the distortions of the reconstructions at the sink, as long as they remain in the shown rate region~\eq{eq:rates_dist}.

We account for the cost of source acquisition and compression by defining a function $\Pcn(R_{n}(t))$ that provides the power spent for compressing the acquired data at a particular rate $R_{n}(t)$. For the sake of analytical tractability, we assume that each function $\Pcn(R_{n}(t))$ is 
\be 
\label{eq:compr_energy}
\Pcn(R_{n}(t)) = \alpha_{n}R_{n}(t), 
\ee
for some coefficient $\alpha_{n} \geq 0$.
Finally, we remark that the destination $d$ is assumed not to have sensing capabilities, and thus is not able to acquire any measurements. We will treat the extension to this setting in Section~\ref{sec:side_information}.

\subsection{Energy Model}
\label{sec:energy_model}
Every node in the network is assumed to be powered via energy harvesting. The harvested energy is stored in an energy storage device, or battery, which is modeled as an energy queue, as in e.g.,~\cite{Huang10}. The energy queue size $E_{n}(t)$ at a node $n \in \N$ measures the amount of energy left in the battery of a node $n$ at the beginning of time slot $t$. For convenience, we normalize the available energy to the number of channel uses (transmission symbols) per slot. Without loss of generality, we assume unitary slot duration so that the amount of power consumed for transmission and acquisition/compression is equivalent to the energy spent in a time slot. Therefore, at any time slot $t$, the overall energy used at a node $n \in \N$ must satisfy the availability constraint
\be
P_n + \Pcn(R_n(t)) \leq E_{n}(t).
\label{eq:constr_energy}
\ee
That is, the total consumed energy due to transmission and acquisition/compression must not exceed the energy available at the node. 

We denote by $H_{n}(t) \leq \Hmax$ the amount of energy harvestable by node $n$ at time slot $t$, and we define the vector $\H(t) = (H_{1}(t), \ldots, H_{N}(t))$ as the {\it energy-harvesting state}. We assume that $\H(t)$ takes value in a finite set $\Hset = \{ \h_{1}, \h_{2}, \ldots, \h_{P}\}$, and is constant for the duration of a time slot, but i.i.d. over time slots. Finally, we define the probability $\rho_{\h_{i}} = \Pr[ \H(t)=\h_{i}]$, which accounts for possible spatial correlation of the harvestable energy across different nodes. 

The energy harvested at time $t$ is assumed to be available for use at time $t+1$. Moreover, each node $n \in \N$ can decide how much of the harvestable energy $H_{n}(t)$ to store in the battery at time slot $t$, and we denote the harvesting decision by $\Htilde_{n}(t)$, with  $0 \leq \Htilde_{n}(t) \leq H_{n}(t)$. We define the harvesting decision vector as $\Htildebf(t) = (\Htilde_{1}(t), \ldots, \Htilde_{N}(t))$. Variable $\Htilde_{n}(t)$ is introduced, following~\cite{Huang10}, to address the issue of assessing the needs of the system in terms of capacities of the energy storage devices. In fact, as in~\cite{Huang10}, we do not make any assumption about the battery maximum size. However, it will be proved later that performance arbitrarily close to the optimal attainable with no limitations on the battery capacity can be achieved with finite-capacity batteries.

\subsection{Queueing Dynamics}

We now detail the dynamics of the network queues. We define $\E(t) = (E_{1}(t), \ldots, E_{N}(t))$ to be the vector of the energy queue sizes of all nodes at time $t$. From the discussion above, for each node $n \in \N$, $E_{n}(t)$ evolves as
\be
E_{n}(t+1) = E_{n}(t) - P_n(t) - \Pcn(R_{n}(t)) + \Htilde_{n}(t),
\label{eq:energy_queue}
\ee
since at each time slot $t$, the energy $P_n(t) + \Pcn(R_{n}(t))$ is consumed, while energy $\Htilde_{n}(t)$ is harvested. We assume $E_{n}(0) \geq 0$ for all $n \in \N$.

We also define the vector $\U(t) = (U_{1}(t), \ldots, U_{N}(t))$, for each time slot $t$, to be the network data queue backlog, where $U_{n}(t)$ represents the amount of data queued at node $n$, which is normalized on the number of channel uses per time slot for convenience of notation, that is it is expressed in terms of bits over channel uses per slot. Denote as $b$ the ratio between the number of channel uses per slot and the number of source samples per slot. Since $b$ typically accounts for the ratio of the channel and source bandwidth, it is conventionally referred to as {\it bandwidth ratio},~\cite{ElGamal12}. We assume that each queue $U_{n}(t)$ evolves according to the following dynamics:
\be
U_{n}(t+1) \leq \max \left \{ U_{n}(t) - \muout(t), 0 \right \} + \muin(t) + \frac{R_{n}(t)}{b},
\label{eq:data_queue}
\ee
since at any time slot $t$, each node $n \in \N$ can transmit, and thus remove from its data queue, at most $\muout(t)$ bits per channel use, while it can receive at most $\muin(t)$ bits per channel use due to transmissions from other nodes and $R_{n}(t)/b$ bits per channel use due to data acquisition/compression. We assume that $U_{n}(0)=0$ for all $n \in \N$. Following standard definitions~\cite{Neely10}, we say that the network is stable if the following condition holds true:
\be
\limsup \frac{1}{T} \sum_{t = 0}^{T-1} \sum_{n \in \N} \Exp [ U_{n}(t) ] < \infty .
\label{eq:constr_stab}
\ee
Notice that the network stability condition~\eq{eq:constr_stab} implies that the data queue $U_{n}(t)$ of each node $n \in \N$ is stable in the sense that $\limsup \frac{1}{T} \sum_{t = 0}^{T-1} \Exp [ U_{n}(t) ] < \infty$.

\setcounter{equation}{18}
\begin{figure*}[t]
\begin{eqnarray}
\label{eq:dual_function_oneState}
 & &\hspace{-0.5cm} \dual_{\o_{i},\s_{j},\h_{k}}({\bm \lambda}^{(\o_{i})}, {\bm \upsilon},{\bm \chi}) = \inf_{\R^{(\o_{i})}, \D^{(\o_{i})}, \P^{(\s_{j})}, \Htildebf^{(\h_{k})}} \Bigg \{ \sum_{n \in \N} V \fn(D_{n}^{(\o_{i})})  + \sum_{m=1}^{2^{N}-1} \lambda_{m}^{(\o_{i})}  \Bigg [ \entropy(\X_{m},\o_{i}) -\log \left( (2 \pi e)^{\vert \X_{m} \vert} \prod_{n \in \X_{m}} D_{n}^{(\o_{i})} \right ) \nonumber \\ 
&& \hspace{-0.5cm} \;\;\;\;\;\; - \sum_{n \in \X_{m}} R_{n}^{(\o_{i})} \Bigg ] + \sum_{n \in \N} \upsilon_{n} \Bigg [ \frac{R_{n}^{(\o_{i})}}{b} + \muin(\P^{(\s_{j})},\s_{j}) 
 - \muout(\P^{(\s_{j})},\s_{j}) \Bigg ] + \sum_{n \in \N} \chi_{n} \left [ P_{n}^{(\s_{j})} + \Pcn(R_{n}^{(\o_{i})}) - \Htilde_{n}^{(\h_{k})}\right ] \Bigg \} 
\end{eqnarray}
\end{figure*}

\setcounter{equation}{14}

\subsection{Optimization Problem}
\label{sec:optimization_problem}
Define $\Theta(t) = \left ( \S(t), \O(t), \H(t), \U(t), \E(t)\right )$ as the state of the network at time slot $t$. A (past-dependent) policy $\pi = \{\pi(t)\text{: } t = 1,2, \ldots \}$ is a collection of mappings between the past and current states $\{\Theta(\tau)\text{: } \tau = 1,\ldots, t \}$ and the current decision $(\R(t),\D(t),\Htildebf(t), \P(t))$ on rates $\R(t)$, distortion levels $\D(t)$, harvested energy $\Htildebf(t)$ and transmission powers $\P(t)$. Moreover, for each node $n \in \N$, let $\fn(D_{n}(t))$ denote the cost incurred by node $n$ when its corresponding  distortion is $D_{n}(t)$. We assume that each function $\fn(D_{n}(t))$ is convex, finite and non-decreasing in the interval $[\Dmin, \Dmax]$. Our objective is to solve the following optimization problem: 
\bea
\label{eq:opt_probl}
\underset{\pi}{\operatorname{minimize}}\ F_{0}^{\pi} = \sum_{n \in \N} F_{n}^{\pi}
\eea
where
\be
\label{eq:av_nodeCost}
F_{n}^{\pi} = \fnpidef,
\ee
subject to the rate-distortion constraints~\eq{eq:rates_dist}, the energy availability constraint~\eq{eq:constr_energy} and network stability constraint~\eq{eq:constr_stab}. Note that~\eq{eq:av_nodeCost} is the per-slot average cost for node $n$.

\section{Lower bound}
\label{sec:optimal_penalty}

In this section, we obtain a lower bound on the optimal network cost $F_{0}^{*}$ of problem~\eq{eq:opt_probl}. This result will be used in Section~\ref{sec:performance} to obtain analytical performance guarantees on our online optimization policy, presented in Section~\ref{sec:proposed_algorithm}. The lower bound is expressed in terms of an optimization problem over parameters $\R^{(\o_{i})} = [R_{1}^{(\o_{i})},\ldots,R_{N}^{(\o_{i})}]$ and $\D^{(\o_{i})} = [D_{1}^{(\o_{i})},\ldots, D_{N}^{(\o_{i})}]$ for all $\o_{i} \in \Oset$, $\P^{(\s_{j})}$ with entries $P_{n,m}^{(\s_{j})}$ for each $(n,m) \in \L$ and for all $\s_{j} \in \Sset$, and $\Htildebf^{(\h_{k})} = [\Htilde_{1}^{(\h_{k})},\ldots,\Htilde_{N}^{(\h_{k})}]$ for all $\h_{k} \in \Hset$. The proof is based on relaxing the stability constraint~\eq{eq:constr_stab} by imposing the necessary condition that the average arrival rate at each data queue be smaller than or equal to the average departure rate, and by also relaxing the energy availability constraint~\eq{eq:constr_energy} by requiring it to be satisfied only on average. Finally, Lagrange relaxation is used on the resulting problem. The details of the proof are available in Appendix~\ref{sec:proof_thm:optimal}.

\thm{
\label{thm:optimal}
The optimal network cost $F_{0}^{*}$ satisfies the following inequality:
\be
V F_{0}^{*} \geq \dual({\bm \lambda}, {\bm \upsilon},{\bm \chi}),
\ee
for all ${\bm \lambda} \in \mathbb{R}^{L(2^{N}-1)}_{+},{\bm \upsilon} \in \mathbb{R}^{N}_{+}, {\bm \chi} \in \mathbb{R}^{N}$, where $\dual({\bm \lambda}, {\bm \upsilon},{\bm \chi})$ is given by
\be
\label{eq:dual_function}
\begin{split}
\dual({\bm \lambda},& {\bm \upsilon},{\bm \chi}) = \sum_{\o_{i} \in \Oset} \rho_{\o_{i}} \! \sum_{\s_{j} \in \Sset} \! \rho_{\s_{j}} \! \sum_{\h_{k} \in \Hset} \! \rho_{\h_{k}} d_{\o_{i},\s_{j},\h_{k}}({\bm \lambda}^{(\o_{i})}, {\bm \upsilon},{\bm \chi}),
\end{split}
\ee

\setcounter{equation}{19}

with $d_{\o_{i},\s_{j},\h_{k}}({\bm \lambda}^{(\o_{i})}, {\bm \upsilon},{\bm \chi})$ defined in \eq{eq:dual_function_oneState}, where the infimum is taken under constraints:
\bea
&& \hspace{-1.5cm}  \begin{split}
0 \leq R_{n}^{(\o_{i})} \leq \Rmax,&\Dmin \leq D_{n}^{(\o_{i})} \leq \Dmax, \\
&\text{for all } n \in \N, \o_{i} \in \Oset, \label{eq:constr_RD}
\end{split}\\
&& \hspace{-1.5cm} 0 \leq P_{n}^{(\s_{j})} \leq \Pmax, \text{for all } n \in \N, \s_{j} \in \Sset, \label{eq:constr_P} \\
&& \hspace{-1.5cm} \text{and }Ê0 \leq \Htilde_{n}^{(\h_{k})} \leq \h_{k,n}, \text{for all } n \in \N, \h_{k} \in \Hset. \label{eq:constr_Htilde} 
\eea
}
\begin{IEEEproof}
See Appendix~\ref{sec:proof_thm:optimal}. 
\end{IEEEproof}

\section{Proposed Policy}
\label{sec:proposed_algorithm}

In this section, we propose an algorithm designed following the Lyapunov optimization framework, as developed in~\cite{Georgiadis06}~\cite{Neely10}, to solve the optimization problem~\eq{eq:opt_probl}. In particular, we aim at finding an online policy $\pi$ for problem~\eq{eq:opt_probl} with close-to-optimal performance, by using Lyapunov optimization with weight perturbation. The technique of weight perturbation, as proposed in~\cite{Huang10}, is used to ensure that the energy queues are kept close to a target value. This is done to avoid battery underflow in a way that is reminiscent of the battery management strategies put forth in~\cite{Srivastava10}, and is further discussed below.

The proposed policy operates by approximately minimizing at each time slot the one-slot conditional Lyapunov drift plus penalty~\cite{Neely10} of the energy and data queues (\eq{eq:energy_queue} and~\eq{eq:data_queue}, respectively) of the network. The optimization is done in an on-line fashion based on the knowledge of the current channel state $\S(t)$, observation state $\O(t)$, data queue sizes $\U(t)$ and energy queue sizes $\E(t)$. Note that no knowledge of the statistics of the states is required, as it is standard with Lyapunov optimization techniques~\cite{Georgiadis06,Neely10}. Using this approach, we obtained the following online optimization algorithm.

{\it Algorithm}: Fix a weight $\thetabf = [\theta_{1}, \ldots, \theta_{N}] \in \mathbb{R}^{N}_{+}$ and a parameter $V >0$. At each time slot $t$, based on the values of the queues $\E(t)$ and $\U(t)$, channel states $\S(t)$ and observation states $\O(t)$, perform the following:
\bi
\e {\it Energy Harvesting:} For each node $n \in \N$, choose $\Htilde_{n}(t)$ that minimizes $(E_{n}(t) - \theta_{n})\Htilde_{n}(t)$ under the constraint $0 \leq \Htilde_{n}(t) \leq H_{n}(t)$. That is, if $(E_{n}(t) - \theta_{n})<0$, perform energy harvesting and store the harvested energy, i.e., set $\Htilde_{n}(t) = \min \{ \theta_{n}-E_{n}(t), H_{n}(t)\}$; otherwise, perform no harvesting, i.e., set $\Htilde_{n}(t) = 0$;
\e {\it Rate-Distortion Optimization:} Choose the source acquisition/compression rate vector $\R(t) = \r = [r_{1}, \ldots, r_{N}]$ and the distortion levels $\D(t) = \d = [d_{1}, \ldots, d_{N}]$ to be an optimal solution of the following optimization problem:
\be
\label{eq:rate_dist_alloc}
\begin{split}
\underset{\r,\d}{\operatorname{minimize\ }} \sum_{n \in \N} [ U_{n}(t)r_{n} - (E_{n}(t)-\theta_{n})\Pcn(r_{n})\\ 
+ V \fn(d_{n}) ],
\end{split}
\ee
subject to the rate-distortion region constraint~\eq{eq:rates_dist}, and to the constraints $0 \leq r_{n} \leq \Rmax$ and $\Dmin \leq d_{n} \leq \Dmax$, for all $n \in \N$;
\e {\it Power Allocation:} Define the weight of a link $(n,m)$ as
\be
\label{eq:link_weight}
W_{n,m}(t) = \max \{ U_{n}(t) - U_{m}(t) - \delta, 0\},
\ee
where $\delta = \lmax \mumax + \Rmax$, and choose $\P(t) = \p$ with entries $p_{n,m}$ for $(n,m) \in \L$ to be an optimal solution of the following optimization problem:
\be
\label{eq:power_alloc}
\begin{split}
\underset{\p}{\operatorname{maximize\ }} \sum_{n \in \N} \Big [ \sum_{m \in \N \setminus n} \Cap_{n,m}(\p,\S(t)) W_{n,m}(t)\\ 
+ (E_{n}(t)-\theta_{n})p_{n} \Big ],
\end{split}
\ee
where $p_{n} = \sum_{m \in \N\setminus n} p_{n,m}$, subject to constraints $0 \leq p_{n} \leq \Pmax$, for each $n \in \N$;
\e {\it Queues Update:} Update $\E(t)$ and $\U(t)$ according to~\eq{eq:energy_queue} and~\eq{eq:data_queue}, respectively.
\ei
\rem{\label{rem:energy_availability} In the algorithm proposed above, the energy availability constraint~\eq{eq:constr_energy} is not explicitly imposed. However, as discussed in Section~\ref{sec:performance}, with a proper choice of the weight vector $\thetabf$, the battery levels are guaranteed to be such that condition~\eq{eq:constr_energy} is never violated. In other words, the effect of the weight vector $\thetabf$ is to ensure that, whenever the algorithm requires to draw energy from the batteries for transmission or acquisition/compression, there is energy available at the corresponding nodes to satisfy the request.
}

\subsection{Price-based Distributed Optimization}
\label{sec:distributed_opt}

While the {\it Energy Harvesting} step can be performed independently by all nodes, the {\it Rate-Distortion Optimization} problem~\eq{eq:rate_dist_alloc} and the {\it Power Allocation} problem~\eq{eq:power_alloc} require centralized optimization. Decentralized implementations of the {\it Power Allocation} problem~\eq{eq:power_alloc} are discussed in many papers, see, e.g.,~\cite{Dohler07}. Here we discuss how to (approximately) solve the {\it Rate-Distortion Optimization} problem~\eq{eq:rate_dist_alloc} in a distributed fashion via dual decomposition~\cite{bertsekasBook}~\cite{Palomar06}. To this end, we introduce the Lagrange multipliers ${\bm \lambda} \in \mathbb{R}^{2^{N}-1}_{+}$ for the $2^{N}-1$ coupling constraints~\eq{eq:rates_dist}, thus obtaining the Lagrangian function for problem~\eq{eq:rate_dist_alloc}:
\begin{eqnarray}
\label{eq:lagr_ratedist}
\hspace{-0.5cm} \Lag(\r,\d,{\bm \lambda})  \! \! & = & \! \! \! \sum_{n \in \N} [ U_{n}(t)r_{n} - (E_{n}(t)-\theta_{n})\Pcn(r_{n}) \nonumber \\  
\hspace{-0.5cm}\! \! &+& \! \! \! V \fn(d_{n})  ] + \sum_{m} \lambda_{m}\Bigg [ \entropy(\X_{m},\O(t)) \nonumber \\ 
\hspace{-0.5cm}\! \! &-& \! \! \! \log \left( (2 \pi e)^{| \X_{m} |} \prod_{l \in \X_{m}} d_{l} \right ) - \sum_{l \in \X_{m}} r_{l} \Bigg ] , 
\end{eqnarray}
where the second sum runs over all the $2^{N}-1$ subsets $\X_{m}$ of $\N$. We will use the notation $\X_{m}$ for the subsets of $\N$ throughout the rest of the paper. Moreover, the dual function for problem~\eq{eq:rate_dist_alloc} is
\be
\label{eq:dualRate_dist_alloc}
\Gfunc({\bm \lambda}) = \inf_{\r, \d} \Lag(\r,\d,{\bm \lambda}),
\ee
with constraints $0 \leq r_{n} \leq \Rmax$ and $\Dmin \leq d_{n} \leq \Dmax$ and the Lagrange dual problem is given by
\be
\label{eq:dual_ratedist}
\underset{{\bm \lambda} \succeq 0}{\operatorname{maximize\ }} \Gfunc({\bm\lambda}).
\ee
Following the dual decomposition approach~\cite{bertsekasBook}~\cite{Palomar06}, the problem of calculating the dual function~\eq{eq:dualRate_dist_alloc} for a given multiplier vector $\bm \lambda$ can be decomposed into $N$ local optimization subproblems, one for each node $n \in \N$. Moreover, solution of the dual problem~\eq{eq:dual_ratedist} can be performed in an iterative fashion using the subgradient method~\cite{bertsekasBook}, as it is standard practice~\cite{bertsekasBook}~\cite{Palomar06}. This leads to the following price-based distributed iterative solution of the dual problem~\eq{eq:dual_ratedist} for time slot $t$: 

Initialize ${\bm \lambda}(1)\succeq 0$. Then, for each iteration $\tau = 1,2, \ldots$:
\bi
\e For the given ${\bm \lambda}(\tau) = {\bm \lambda}$, each source node $n$ solves the local optimization problem
\begin{eqnarray}
\label{eq:local_optim}
& & \hspace{-1.5cm} \underset{0 \leq r_{n} \leq \Rmax,\ \Dmin \leq d_{n} \leq \Dmax}{\operatorname{minimize\ }} U_{n}(t)r_{n} - (E_{n}(t)-\theta_{n})\Pcn(r_{n}) \nonumber \\
& + & V\fn(d_{n}) - (\log(d_{n}) + r_{n})\sum_{m\text{: }n \in \X_{m}} \lambda_{m},
\end{eqnarray}
obtaining the optimal values $(r_{n}^{*}({\bm \lambda}),d_{n}^{*}({\bm \lambda}))$;
\e The dual variables ${\bm \lambda}$ are updated using the subgradient method~\cite[Section 6.1]{bertsekasBook} as
\be
\label{eq:lambda_tau}
{\bm \lambda}(\tau + 1) = {\bm \lambda}(\tau) + \epsilon_{\tau}a({\bm \lambda(\tau)}),
\ee
where $\epsilon_{\tau}$ is a positive scalar step size and $a({\bm \lambda}) = \sum_{m} \entropy(\X_{m},\O(t)) -\log (2 \pi e)^{| \X_{m} |} - \sum_{n \in \N}  \log(d_{n}^{*}({\bm \lambda}))+r_{n}^{*}({\bm \lambda})$ is a subgradient of function $\Gfunc({\bm\lambda})$.
\ei
With various choices for the weights $\epsilon_{\tau}$ (e.g., $\epsilon_{\tau} = 1/\tau$), due to the concavity of function $\Gfunc({\bm\lambda})$, the procedure above is guaranteed to converge to the optimal value of the dual problem~\eq{eq:dual_ratedist}~\cite[Section 3.4]{bertsekasBook}. Moreover, under the given assumptions, problem~\eq{eq:rate_dist_alloc} is convex and satisfies Slater's condition~\cite{boyd2004}. Therefore, strong duality holds, which guarantees that the optimal value of the dual problem~\eq{eq:dual_ratedist} coincides with the optimal value of~\eq{eq:rate_dist_alloc}, and the optimal value of~\eq{eq:dual_ratedist} is attained at some value ${\bm \lambda}^{*}$. However, in order for the illustrated iterative procedure to converge to an optimal solution $(\r^{*},\d^{*})$ of problem~\eq{eq:rate_dist_alloc}, we need that the value of the pair\footnote{This pair exists in virtue of the Weierstrass theorem~\cite{bertsekasBook}.} $(\r,\d)$ at which the infimum in~\eq{eq:dualRate_dist_alloc} is attained for ${\bm \lambda} = {\bm \lambda}^{*}$ coincides with the optimal pair for the original problem~\eq{eq:rate_dist_alloc}. This can be guaranteed if the Lagrangian function $\Lag(\r,\d,{\bm \lambda})$ is strictly convex in $(\r,\d)$~\cite[Section 3.4]{bertsekasBook}. As proposed in~\cite{Xiao04} this can be enforced by adding a small term $\epsilon(||\r||^{2}+||\d||^{2} )$ to $\Lag(\r,\d,{\bm \lambda})$ while performing the minimization~\eq{eq:dualRate_dist_alloc}, foe some $\epsilon >0$. Although this operation is bound to make the solution only approximate, the quality of the approximation can be controlled by keeping $\epsilon$ small.

\section{Performance Analysis}
\label{sec:performance}

In this section, we provide analytical insights into the performance of the policy proposed in Section~\ref{sec:proposed_algorithm}. To this end, we define the parameters $\beta_{n} = \min \left \{ \alpha_{n}, 1 \right \}$ (recall~\eq{eq:compr_energy}) and $\gamma_{n} = \sup_{\Dmin \leq d_{n} \leq \Dmax} \left [ \frac{\fn(d_{n})-\fn(\Dmax)}{\log(d_{n}/\Dmax)} \right ]$, which is finite under the given assumptions.

\thm{ 
\label{thm:performance}
Under the proposed algorithm with $\thetabf = [\theta_{1}, \ldots, \theta_{N}]$, where $\theta_{n} = \frac {\gamma_{n}}{\beta_{n}} V + \alpha_{n} \Rmax + \Pmax$, we have:
\ben
\e The data queue and the energy queue of all nodes are bounded as:
\bea
\label{eq:energy_bound}
0 \leq &E_{n}(t)& \leq \theta_{n},\\
\text{and } 0 \leq &U_{n}(t)& \leq  \gamma_{n} V  + \Rmax ,
\label{eq:data_bound}
\eea
respectively, for all nodes $n \in \N$ and all times $t$;
\e When a node $n \in \N$ allocates a non-zero power to any of its outgoing links (i.e., $P_{n}(t) > 0$), and/or when it chooses a non-zero source acquisition rate (i.e., $R_{n}(t)>0$), thus expending energy for source acquisition/compression, we have that:
\be
\label{eq:enough_energy}
E_{n}(t) \geq \alpha_{n}\Rmax + \Pmax.
\ee
This condition guarantees that the energy availability constraint~\eq{eq:constr_energy} is satisfied for all nodes $n \in \N$ and all times $t$ (see Remark~\ref{rem:energy_availability} and Remark~\ref{rem:energy_availability2}).
\e The overall cost $F^{\pi}_{0}$ \eq{eq:opt_probl} achieved by the proposed scheme satisfies the bound
\be
F^{\pi}_{0} = \sum_{n \in \N} F^{\pi}_{n} \leq F_{0}^{*} + \frac{B}{V},
\ee
where $F_{0}^{*}$ is the optimal cost of problem~\eq{eq:opt_probl} and the finite constant $B$ is $B = N \left ( \mumax(\mumax + \Rmax) + \RmaxSq/2 \right )+ N/2 ( \HmaxSq +\PcmaxSq + \PmaxSq+2\Pcmax\Pmax ) + N (\delta \lmax \mumax +  \HmaxSq/4)$.
\een
}

\begin{IEEEproof}
See Appendix~\ref{sec:proof_thm:performance}.
\end{IEEEproof}

\rem{\label{rem:energy_availability2} The fact that~\eq{eq:enough_energy} implies that the proposed algorithm satisfies the energy availability constraint~\eq{eq:constr_energy} at each time slot follows since each node $n \in \N$ cannot consume an energy larger that $\alpha_{n}\Rmax + \Pmax$ in a time slot. In fact, $\alpha_{n}\Rmax$ is the maximum energy spent for compressing the acquired data and $\Pmax$ is the maximum transmission energy consumption.
}

\rem{\label{rem:koksal_paper} Following~\cite{Mao11}, under the modified stability requirement
$\limsup \frac{1}{T} \sum_{t = 1}^{T-1} U_{n}(t) < \infty, \text{ for all } n \in \N,$ the proposed algorithm can be proven to guarantee near-optimal performance with probability one.
}

\section{Extension with side information at the sink}
\label{sec:side_information}
We now consider an extended version of the problem studied thus far, in which the sink node $d$, rather than being the final destination for the sources measured at the sensors, acts as a cluster head and communicates to a network collector node $c$ (see Fig.~\ref{fig:scenario}), on a communication link modeled as for any other pair of node (see Section~\ref{sec:transmission_model}). The key novel aspect of this extended model is that node $d$ can measure a source correlated with that of the sensors and use such side information to improve the system performance. Specifically, thanks to the side information available at node $d$, the rate requirements for communication from the sensors to $d$ can be reduced. However, node $d$, which is powered by energy-harvesting as all the sensors, also needs to communicate with node $c$. Therefore, a new trade-off arises between the energy allocated by $d$ to acquire side information and that used by $d$ to communicate with $c$.  

We now discuss how the model discussed in Section~\ref{sec:system_model} needs to be modified in order to account for the different setting of interest here. First, the destination $d$ acquires a source signal which is correlated with the sensor's measures with a rate $R_{d}(t)$. This affects the rate-distortion constraints~\eq{eq:rates_dist} in that the entropy function $\entropy(\X,\O(t))$ should now be conditioned on the side information available at the receiver (see, e.g.,~\cite{Gastpar04}). This leads to modified rate-distortion constraints~\eq{eq:rates_dist} with a function $\entropy(\X,\O(t),R_{d}(t))$ that also depends on $R_{d}(t)$. An example of this function will be given in Section~\ref{sec:results}. The energy used for acquiring the side information is given by $\Pcd(R_{d}(t)) = \alpha_{d} R_{d}(t)$ and the slot duration similar to all other nodes. Moreover, the data queue at node $d$ evolves as 
\be
U_{d}(t+1) \leq \max \left \{ U_{d}(t) - \mu_{d,c}(t), 0 \right \} + \mu_{*,d}(t),
\label{eq:data_queue_d}
\ee
where $\mu_{d,c}(t)$ and $\mu_{*,d}(t)$ represent, respectively, the transmitted and received data at time $t$, and transmission is to the collector node $c$. Note that no other node is connected to the network collector $c$ apart from $d$. The energy queue $E_{d}(t)$, instead, evolves according to~\eq{eq:energy_queue}. Finally, $\P(t)$ and $\S(t)$ are extended to consider the additional link $(d,c) \in \L$ and the rate achievable on that link is given by $\Cap_{d,c}(\P(t),\S(t))$, which is assumed to have the same properties as for all other links (see Section~\ref{sec:system_model}). We refer to the power used for transmission by node $d$ as $P_{d}$.

In what follow we modify the algorithm proposed in Section~\ref{sec:proposed_algorithm} in order to address the new setting outlined above. The modified algorithm works as follows:
\bi
\e {\it Energy Harvesting:} Follow the same procedure as for the algorithm discussed in Section~\ref{sec:proposed_algorithm}, for all nodes including node $d$;
\e {\it Rate-Distortion Allocation:} Choose $R_{n}(t)$ and $D_{n}(t)$, $n = 1, \ldots, N$, and $R_{d}(t)$ to be the optimal solution of the following optimization problem:
\be
\begin{split}
\underset{(\r,\d),r_{d}}{\operatorname{minimize}} &\sum_{n \in \N} [ U_{n}(t)r_{n} - (E_{n}(t)-\theta_{n})\Pcn(r_{n}) \\
&+ V \fn(d_{n}) ]+ (E_{d}(t)-\theta_{d})\Pcd(r_{d}),
\end{split}
\label{eq:side_info_opt}
\ee
subject to $\sum_{n \in \X} r_{n} \geq \entropy(\X,\O(t),r_{d}) - \log \left ( (2 \pi e)^{\vert \X \vert} \prod_{n \in \X} d_{n} \right ), \forall \X \subseteq \N$,  $0 \leq r_{n} \leq \Rmax$ and $\Dmin \leq d_{n} \leq \Dmax$, $n \in \N$ and $0 \leq r_{d} \leq \Rmax$;
\e {\it Power Allocation:} Define the weight of a link $(n,m) \in \L$ as\footnote{We remind that $\L$ is extended to consider the link $(d,c)$.}~\eq{eq:link_weight} and choose $\P(t) = \p$ with entries $p_{n,m}$ for $(n,m) \in \L$ to be an optimal solution of the following optimization problem:
\be
\label{eq:power_alloc_side}
\begin{split}
&\underset{\p}{\operatorname{maximize\ }} \sum_{n \in \N} \Big [ \sum_{m \in \N \setminus n} \Cap_{n,m}(\p,\S(t)) W_{n,m}(t)\\ 
&+ (E_{n}(t)-\theta_{n})p_{n} \Big ]+\Cap_{d,c}(\p,\S(t)) W_{d,c}(t)\\
&+ (E_{d}(t)-\theta_{d})p_{d},
\end{split}
\ee
subject to $0 \leq p_{n} \leq \Pmax$, for each $n \in \N \cup \{ d\}$.
\e {\it Queues Update:} Update $\E(t)$ and $E_{d}(t)$ according to~\eq{eq:energy_queue}, $\U(t)$ according to~\eq{eq:data_queue} and $U_{d}(t)$ using~\eq{eq:data_queue_d}.
\ei

The algorithm proposed above is a simple modification of the algorithm proposed in Section~\ref{sec:proposed_algorithm} that accounts for the need to allocate rate and power also for node $d$. It can be proven that this algorithm has similar optimality properties as the algorithm of Section~\ref{sec:proposed_algorithm} as summarized in Theorem~\ref{thm:performance}. We omit a formal statement of this result here, since it is a straightforward extension of Theorem~\ref{thm:performance}.

\section{Numerical Results}
\label{sec:results}

In this Section, we provide further insights into the performance of the system under study, via some numerical results. We consider the network topology of Fig.~\ref{fig:scenario}, where the set $\N$ of nodes gathers spatially correlated data and transmits it to the sink node $d$. We first consider the set-up without side information at the sink described in Section~\ref{sec:system_model}. We assume that nodes $\{1,2,3\}$ collect the measurements, while nodes $\{4,5\}$ are only used as relays (or equivalently measure zero-power sources). The signal samples measured at nodes $\{1,2,3\}$ are jointly Gaussian with zero mean and time-independent correlation matrix 
\be
\label{eq:corrMatrix}
\O(t) = \left[ \begin{array}{ccc}
1 & \omega & \omega \\
\omega & 1 & \omega \\
\omega & \omega & 1 \end{array} \right],
\ee
where $\omega \in [-1,1]$ is the spatial correlation coefficient. The channel state matrix $\S(t)$ has independent entries that are Rayleigh distributed, while the energy harvesting vector $\H(t)$ has independent entries that are uniformly distributed in $[0, \Hmax]$. Both channel and energy harvesting statistics are i.i.d. across time slots.

For the channel capacity function,  we consider $\Cap_{n,m}(\P(t),\S(t)) = \log (1 + P_{n,m}(t)S_{n,m}(t))$ for all $(n,m) \in \L$, while the entropy function is given by~\eq{eq:entropy_Gauss} and the cost function is $\fn(D_{n}(t))= D_{n}(t)$ for all $n \in \N$. Moreover, we set the numerical values $\alpha_{n}=1$, $\Hmax = 3$, $\Dmin = 0.001$ and $\Pmax = \Pcmax$, with $\Rmax = \entropy(\{1,2,3\},\O(t)) - \log \left ( (2 \pi e \Dmin)^{3} \right )$. In what follows, we refer to {\it network queue size} as the sum of the queue sizes of all nodes in $\N$.

\begin{figure}[t!]
\begin{center}
\includegraphics[width=\figw]{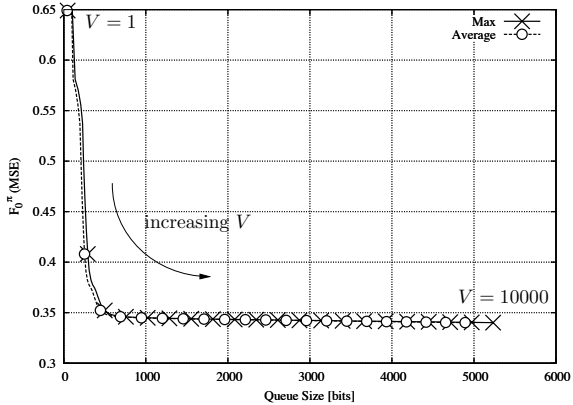}
\caption{$F^{\pi}_{0}$ {\it vs} maximum and average network queue size for $1 \leq V \leq 10000$. Each pair of values for sum-distortion and queue size is obtained for a different value of $V$, from $1$ to $10000$, with step length $500$. ($\omega = 0.5$)}
\label{fig:performance_V}
\end{center}
\end{figure}

We first examine the effect of parameter $V$, which was shown in Theorem~\ref{thm:performance} to characterize the $(V,1/V)$ trade-off between the network queue size and the additive gap with respect to the lower bound of Theorem~\ref{thm:optimal}. To this end, in Fig.~\ref{fig:performance_V}, we set $\omega = 0.5$ and plot the average sum-distortion $F^{\pi}_{0}$ as a function of the maximum and average network queue size for different value of the parameter $V$. Confirming the results of Theorem~\ref{thm:performance}, we observe that the sum-distortion $F^{\pi}_{0}$ gradually converges to the lower bound set by the optimal value $F_{0}^{*}$ for increasing $V$. A closer inspection of the results also reveals an almost linear increase of the maximum and time average network queue size with respect to $V$, as suggested by Theorem~\ref{thm:performance} (not shown). 

\begin{figure}[t!]
\begin{center}
\includegraphics[width=\figw]{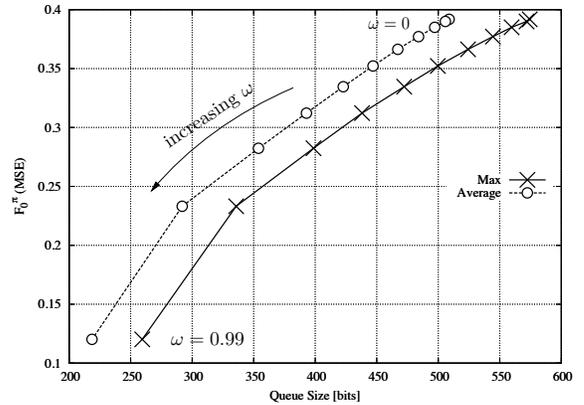}
\caption{$F^{\pi}_{0}$ {\it vs} maximum and average network queue size for different values of source correlation $\omega$, from $0$ to $0.99$, with step length $0.1$. ($V=1000$)}
\label{fig:performance_omega}
\end{center}
\end{figure}

Next, we evaluate the impact of the spatial correlation parameter $\omega$. As discussed, an increasing $\omega$ is expected to lead to a reduction in the energy consumption for the same reconstruction accuracy at the sink thanks to the spatial energy trade-offs enabled by distributed source coding. This is confirmed by the results in Fig.~\ref{fig:performance_omega}, where we plot the sum-distortion $F_{0}^{\pi}$ versus the average and maximum network queue size, where each point is obtained for a different value of the correlation $\omega$ in $[0,1)$. We note that an increasing $\omega$ leads to a reduction of both the network queue size and $F_{0}^{\pi}$. 
Note that the performance with $\omega = 0$ corresponds to that of a conventional source coding system (i.e., not leveraging distributed source coding) as, in this case, distributed source coding does not offer any advantage and reduces to conventional compression. Thus, comparison between the performance with $\omega=0$ and $\omega > 0$ reveals the gain of leveraging distributed source coding. Note that this gain is quite substantial, leading in the best case ($\omega \to 1$) to a decrease of a factor $3$ in terms of distortion and of a factor $2.3$ in terms of queue size at the nodes.  

Finally, we evaluate the performance in the scenario of Section~\ref{sec:side_information}, where the sink node $d$ acts as a cluster head, that measures a source correlated with that of the remaining sensors and communicates the gathered data to node $c$ (see Fig.~\ref{fig:scenario}). To this end, we replace the entropy function $\entropy(\X,\O(t))$ with a function $\entropy(\X,\O(t),R_{d}(t))$, that takes into account the side information obtained by $d$ with rate $R_{d}(t)$. 
We recall that $R_{d}(t)$ is a decision variable of the new problem, see~\eq{eq:side_info_opt}. Following~\cite{Gastpar04}, the function $\entropy(\X,\O(t),R_{d}(t))$ is given by~\eq{eq:entropy_Gauss} where the correlation matrix~\eq{eq:corrMatrix} should now be conditioned on the side information available at the destination~\cite{Gastpar04}. According to the simple source model described in Appendix~\ref{sec:source_model}, we assume that this conditional covariance matrix is given by 
\be
\label{eq:corrMatrixSide}
\!\! \O(t) \! = \! \left[ \!\! \begin{array}{ccc}
1-\omega\omega_{d}(t) & \omega(1-\omega_{d}(t)) & \omega(1-\omega_{d}(t)) \\
\omega(1-\omega_{d}(t)) & 1-\omega\omega_{d}(t) & \omega(1-\omega_{d}(t)) \\
\omega(1-\omega_{d}(t)) & \omega(1-\omega_{d}(t)) & 1-\omega\omega_{d}(t) \end{array} \!\! \right],
\ee
where $\omega_{d}(t) = 1 - 2^{-R_{d}(t)}$. We consider the same simulation parameters as above and we additionally set $\alpha_{d} = 1$ and, only for node $d$, $\Hmax = 12$. 

\begin{figure}[t!]
\begin{center}
\includegraphics[width=\figw]{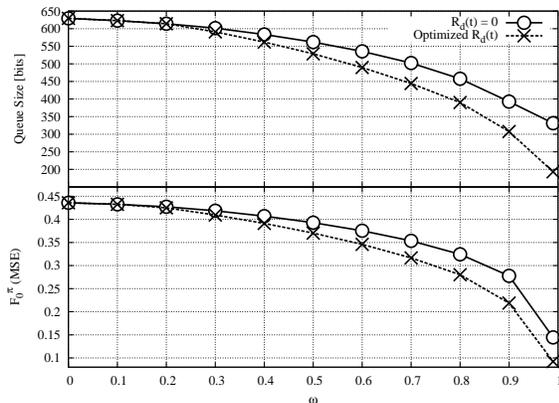}
\caption{$F^{\pi}_{0}$ and average network queue size {\it vs} source correlation $\omega$. ($V=1000$)}
\label{fig:performance_omegaSIDE}
\end{center}
\end{figure}

Fig.~\ref{fig:performance_omegaSIDE} shows the sum-distortion $F_{0}^{\pi}$ and the average network queue size versus $\omega \in [0,1)$. As a reference, we compare the performance of the proposed algorithm with that of a scheme that sets $R_{d}(t)=0$. This scheme, therefore, does not acquire side information at the sink and instead utilizes all the available energy at the sink for transmission to node $c$. It can be seen that gains in terms of memory and distortion can be obtained by properly allocating the available energy between the tasks of transmission and source coding at the sink node, e.g., a reduction of more than $25$\% and $21$\% is obtained for $\omega \geq 0.9$ for the queue size and $F_{0}^{\pi}$, respectively. 

\section{Conclusions}
\label{sec:conclusions}
Energy harvesting poses new challenges in terms of energy management of wireless networks. In sensor networks, these challenges are compounded by the need for balancing the energy consumed by source coding tasks (i.e., data compression) against  that used for transmission. Moreover, the correlation among the data readings collected by different sensors, if leveraged via distributed source coding, makes it possible to exploit spatial energy trade-offs across the sensors, thus allowing for better performance in terms of memory usage and distortion at the sink. Based on the above, this work has proposed a dynamic online optimization strategy for multihop wireless sensor networks with energy harvesting capabilities. This strategy jointly optimizes source coding and data transmission activities for time-varying sources and channels, by ensuring queue stability at the nodes and energy neutrality. The proposed technique, based on Lyapunov optimization, has been analytically shown to be characterized by a $(V,1/V)$ trade-off, revealing a linear relationship for queue and battery size ($V$) and an inverse proportionality in terms of optimality gap ($1/V$), where $V$ is a tunable parameter of the algorithm. Numerical results have demonstrated the key role of source correlation and distributed source coding in the system performance.

\bibliographystyle{IEEEtran}
\bibliography{IEEEabrv,paper}

\appendices

\section{Proof of Theorem~\ref{thm:optimal}}
\label{sec:proof_thm:optimal}
\begin{IEEEproof}
Define as $\phi^{*}$ the optimal value of the following problem:
\be
\label{eq:primal_problem}
\operatorname{minimize} \ V\sum_{n \in \N} \sum_{\o_{i} \in \Oset} \rho_{\o_{i}} \sum_{k=1}^{K} \vartheta_{k}^{(\o_{i})} \fn \left ( D_{n,[k]}^{(\o_{i})} \right )
\ee
subject to:
\bea
&& \hspace{-0.75cm} \entropy(\X,\o_{i}) - \log(2 \pi e)^{\vert \X \vert} \prod_{n \in \X} D_{n,[k]}^{(\o_{i})} \nonumber \\ 
&& \hspace{-0.75cm} \;\; \leq \sum_{n \in \X} R_{n,[k]}^{(\o_{i})}, \text{for all } \X \subseteq \N,\o_{i} \in \Oset, k \in [1, \ldots, K], \label{eq:rate_dist_constr_primal} \\
&& \hspace{-0.75cm} \sum_{\o_{i} \in \Oset}\rho_{\o_{i}}\sum_{k=1}^{K}\vartheta_{k}^{(\o_{i})}\frac{R_{n,[k]}^{(\o_{i})}}{b} + \sum_{\s_{i} \in \Sset}\rho_{\s_{i}}\sum_{k=1}^{K}\varrho_{k}^{(\s_{i})}\muin(\P_{k}^{(\s_{i})},\s_{i}) \nonumber \\ 
&& \hspace{-0.75cm} \;\; \leq \sum_{\s_{i} \in \Sset}\rho_{\s_{i}}\sum_{k=1}^{K}\varrho_{k}^{(\s_{i})}\muout(\P_{k}^{(\s_{i})},\s_{i}), \text{for all } n \in \N, \label{eq:data_queue_constr_primal}\\
&& \hspace{-0.75cm} \sum_{\s_{i} \in \Sset}\rho_{\s_{i}}\sum_{k=1}^{K}\varrho_{k}^{(\s_{i})}\left ( P_{n,[k]}^{(\s_{i})} \right ) + \sum_{\o_{i} \in \Oset}\rho_{\o_{i}}\sum_{k=1}^{K}\vartheta_{k}^{(\o_{i})}\Pcn\left (R_{n,[k]}^{(\o_{i})} \right ) \nonumber \\
&& \hspace{-0.75cm} \;\; = \sum_{\h_{i} \in \Hset}\rho_{\h_{i}}\sum_{k=1}^{K}\varphi_{k}^{(\h_{i})}\Htilde_{n,[k]}^{(\h_{i})}, \text{for all } n \in \N, \label{eq:energy_queue_constr_primal} \\
&& \hspace{-0.75cm} 0 \leq \vartheta_{k}^{(\o_{i})}, \varrho_{k}^{(\s_{i})},\varphi^{(\h_{i})}_{k} \leq 1, \nonumber\\ 
&& \hspace{-0.75cm} \;\;\;\;\;\;\;\; \text{for all } \o_{i} \in \Oset, \s_{i} \in \Sset, \h_{i} \in \Hset, k \in [1, \ldots, K], \nonumber \\
&& \hspace{-0.75cm} \sum_{k=1}^{K}\vartheta_{k}^{(\o_{i})} = 1, \sum_{k=1}^{K}\varrho_{k}^{(\s_{i})} = 1, \sum_{k=1}^{K}\varphi^{(\h_{i})}_{k} = 1,\nonumber \\ 
&& \hspace{-0.75cm} \;\;\;\;\;\;\;\; \text{for all } \o_{i} \in \Oset, \s_{i} \in \Sset, \h_{i} \in \Hset, \nonumber \\
&& \hspace{-0.75cm} 0 \leq R_{n,[k]}^{(\o_{i})} \leq \Rmax, \Dmin \leq D_{n,[k]}^{(\o_{i})} \leq \Dmax, \nonumber \\ 
&& \hspace{-0.75cm} \;\;\;\;\;\;\;\; \text{for all } n \in \N, \o_{i} \in \Oset, k \in [1, \ldots, K], \nonumber \\
&& \hspace{-0.75cm} 0 \leq P_{n,[k]}^{(\s_{i})} \leq \Pmax, \text{for all } n \in \N, \s_{i} \in \Sset, k \in [1, \ldots, K], \nonumber \\
&& \hspace{-0.75cm} 0 \leq \Htilde_{n,[k]}^{(\h_{i})} \leq \h_{i,n}, \text{for all } n \in \N, \h_{i} \in \Hset, k \in [1, \ldots, K], \nonumber
\eea
where the minimization is taken over variables $\vartheta_{k}^{(\o_{i})}$, $\varrho_{k}^{(\s_{i})}$, $\varphi_{k}^{(\h_{i})}$, $R_{n,[k]}^{(\o_{i})}$, $D_{n,[k]}^{(\o_{i})}$, $\Htilde_{n,[k]}^{(\h_{i})}$ and $P_{n,[k]}^{(\s_{i})}$ for all $n \in \N$, $\o_{i} \in \Oset$, $\s_{i} \in \Sset$, $\h_{i} \in \Hset$ and $k \in [1, \ldots, K]$, with $K = 2N +2$. Variables $\left \{R_{n,[k]}^{(\o_{i})} \right \}_{k = 1}^{K}$ and $\left \{D_{n,[k]}^{(\o_{i})} \right \}_{k = 1}^{K}$ can be interpreted, respectively, as the set of rates and distortions selected by node $n \in \N$ when the source state is $\O(t)=\o_{i}$. Specifically, node $n$ selects rate $R_{n,[k]}^{(\o_{i})}$ and distortion $D_{n,[k]}^{(\o_{i})}$ with probability $\vartheta_{k}^{(\o_{i})}$ when the source state is $\O(t)=\o_{i}$. Variables $\left \{ P_{n,m,[k]}^{(\s_{i})} \right \}_{k=1}^{K}$ can be seen as the transmission powers allocated to link $(n,m) \in \L$, when the channel state $\S(t) = \s_{i}$. Each power $P_{n,m,[k]}^{(\s_{i})}$ is selected with probability $\varrho_{k}^{(\s_{i})}$ if $\S(t) = \s_{i}$. Finally, variables $\left \{\Htilde_{n,[k]}^{(\h_{i})} \right \}_{k = 1}^{K}$ represent the harvested energy when the energy harvesting state is $\H(t) = \h_{i} = [h_{i,1},\ldots, h_{i,N}]$. Each energy $\Htilde_{n,[k]}^{(\h_{i})}$ is selected with probability $\varphi_{k}^{(\h_{i})}$ if $\H(t)=\h_{i}$. Note that we added the constant $V$ in the optimization function for our later analysis.

\thm{
\label{thm:optimal_appendix}
The optimal network cost $F_{0}^{*}$ satisfies the following inequality:
\be
V F_{0}^{*} \geq \phi^{*},
\ee
where $\phi^{*}$ is the optimal value of the optimization problem~\eq{eq:primal_problem}.
}
The proof of Theorem~\ref{thm:optimal_appendix} is in Appendix~\ref{sec:proof_thm:optimal_appendix}. 

A generally looser lower bound can be evaluated by the weak duality in Lagrange optimization theory~\cite{boyd2004}, which is easily seen to lead to Theorem~\ref{thm:optimal}. In fact, in~\eq{eq:dual_function}, the parameters $\lambda_{m}^{(\o_{i})}$ for $m = [1, \ldots, 2^{N}-1]$ and $\o_{i} \in \Oset$ are the $L(2^{N}-1)$ Lagrange multipliers corresponding to constraints~\eq{eq:rate_dist_constr_primal}, parameters $\upsilon_{n}$ for $n = [1, \ldots,N]$ are the Lagrange multipliers corresponding to constraints~\eq{eq:data_queue_constr_primal} and parameters $\chi_{n}$ for $n = [1, \ldots,N]$, are the Lagrange multipliers corresponding to constraints~\eq{eq:energy_queue_constr_primal}.
\end{IEEEproof}

\section{Proof of Theorem~\ref{thm:optimal_appendix}}
\label{sec:proof_thm:optimal_appendix}
\begin{IEEEproof}
We follow an argument similar to the one used in~\cite{Huang11}. Consider any stable policy $\pi$, i.e., a policy such that the condition~\eq{eq:constr_stab} is satisfied under $\pi$. Since $\Exp [\muin(t) + R_{n}(t)/b - \muout(t)] \leq (N -1)\mumax + \Rmax/b$, from~\cite[Theorem 2.8]{Neely10}, constraint~\eq{eq:constr_stab} implies the mean rate stability constraint and thus the condition
\be
\begin{split}
\label{eq:mean_rate_stab}
\limsup \frac{1}{T} \sum_{t=0}^{T-1} \Exp &\left [ \muin(t) + \frac{R_{n}(t)}{b} \right ] \leq \\ 
&\liminf \frac{1}{T}  \sum_{t=0}^{T-1} \Exp[ \muout(t)],
\end{split}
\ee
for each node $n \in \N$. We thus relax problem~\eq{eq:primal_problem} by substituting~\eq{eq:constr_stab} with~\eq{eq:mean_rate_stab}. We further relax the energy availability constraint~\eq{eq:constr_energy}, imposing average stability for the energy queues (see \eq{eq:energy_queue})
\be
\begin{split}
\limsup \frac{1}{T} \sum_{t=0}^{T-1} \Exp [P_{n}(t) +& \Pcn(R_{n}(t))] \\
& = \limsup \frac{1}{T} \sum_{t=0}^{T-1} \Exp [ \Htilde_{n}(t)].
\end{split}
\ee
For the relaxed problem, we can show as in~\cite{Huang11} that the optimal policy is stationary and depends only on the source and channel state. From this, by Caratheodory's theorem~\cite{EgglestonBook}, we obtain that the problem at hand is equivalent to~\eq{eq:primal_problem}.
\end{IEEEproof}

\section{Proof of Theorem~\ref{thm:performance}}
\label{sec:proof_thm:performance}
\begin{IEEEproof}

\noindent 1) From the energy harvesting part of the algorithm, we have that $E_{n}(t) \leq \theta_{n}$, since harvesting is performed only when $E_{n}(t) < \theta_{n}$ and the maximum amount of harvested energy in that case is $\theta_{n}-E_{n}(t)$. This proves~\eq{eq:energy_bound}. 
We now prove~\eq{eq:data_bound} by induction on $t$. Inequality~\eq{eq:data_bound} holds for $t=0$, since $U_{n}(0) = 0$ for all $n$. Then, assuming that~\eq{eq:data_bound} is satisfied for all $n$ at time $t$, we show that it holds also for time $t+1$. To this end, we consider separately the different possible cases in which a node $n$ receives or not data from other nodes (i.e., endogenous data) and/or acquires or not its measurement (i.e., exogenous data). First, if node $n$ receives neither endogenous nor exogenous data, then we have that $U_{n}(t+1) \leq U_{n}(t) \leq \gamma_{n} V  + \Rmax$, which proves the claim. Second, assume that node $n \in \N$ receives endogenous, but not exogenous, data. It follows from~\eq{eq:power_alloc} that, for some node $m \in \N$, with $m \neq n$, we must have
\be
\label{eq:cond_endo}
U_{n}(t) \leq U_{m}(t) - \delta \leq \gamma_{n} V  + \Rmax - \delta .
\ee
However, since any node can receive at most $\lmax \mumax$ bits per channel use of endogenous data, we have from~\eq{eq:cond_endo} and the definition of $\delta$ that $U_{n}(t+1) \leq \gamma_{n} V \leq \gamma_{n} V + \Rmax$, which proves the claim. 

We now analyze the case where node $n$ receives exogenous, but not endogenous, data. This implies that $r_{n} > 0$ is obtained from the solution of problem~\eq{eq:rate_dist_alloc}. We define the corresponding Lagrangian function as
\begin{eqnarray}
\label{eq:lagr_ratedist2}
&& \hspace{-1cm} \Lag(\r,\d,{\bm \lambda},{\bm \upsilon}) = \nonumber \\
&& \hspace{-1cm} \;\;\;\; = \sum_{n \in \N} [ U_{n}(t)r_{n} - (E_{n}(t)-\theta_{n})\Pcn(r_{n})+V \fn(d_{n})] \nonumber \\
&& \hspace{-1cm} \;\;\;\;\;\;\;\; + \sum_{m} \lambda_{m} \Big [ \entropy(\X_{m},\O(t)) -\log \Big( (2 \pi e)^{| \X_{m} |} \prod_{l \in \X_{m}} d_{l} \Big ) \nonumber \\ 
&& \hspace{-1cm} \;\;\;\;\;\;\;\; - \sum_{l \in \X_{m}} r_{l} \Big ]  + \sum_{n \in \N} \upsilon_{n} (d_{n} - \Dmax), 
\end{eqnarray}
where we have relaxed the constraints~\eq{eq:rates_dist} and constraints $d_{n} \leq \Dmax$. The Lagrange dual function is given by
\be
\label{eq:dualRate_dist_alloc2}
\Gfunc({\bm \lambda},{\bm \upsilon}) = \inf_{\r, \d} \Lag(\r,\d,{\bm \lambda},{\bm \upsilon}),
\ee
where the infimum is taken with the constraints $0 \leq r_{n} \leq \Rmax$ and $d_{n} \geq 0$, and the dual problem is given by:
\be
\label{eq:dual_ratedist2}
\underset{{\bm \lambda} \succeq 0,{\bm \upsilon} \succeq 0}{\operatorname{maximize\ }} \Gfunc({\bm \lambda},{\bm \upsilon}).
\ee

\lem{	\label{lem:bound_lambda} Any dual optimal vector ${\bm \lambda}^{*}$ (i.e., a vector $\bm \lambda$ maximizing~\eq{eq:dual_ratedist2}) satisfies the conditions
\be
\label{eq:bound_lambda}
\sum_{m\text{: }n \in \X_{m}} \lambda_{m}^{*} \leq \gamma_{n} V,
\ee
for all $n \in \N$. Moreover, any primal optimal $r_{n}^{*}$ satisfies the condition
\be
\label{eq:rn_star}
\begin{split}
r_{n}^{*} = \underset{{0 \leq r_{n} \leq \Rmax}}{\operatorname{argmin\ }} U_{n}(t)r_{n} &- (E_{n}(t)-\theta_{n})\Pcn(r_{n})\\ 
&- r_{n} \sum_{m\text{: }n \in \X_{m}}\lambda_{m}^{*}.
\end{split}
\ee
}

The proof of Lemma~\ref{lem:bound_lambda} can be found in Appendix~\ref{sec:proof_lem:bound_lambda}.

\indent According to~\eq{eq:rn_star} we have that $r_{n}^{*} > 0$ is an optimal solution of problem~\eq{eq:rate_dist_alloc} only if the value of the right-hand side of~\eq{eq:rn_star} evaluated at $r_{n}=0$ is larger than the value obtained by evaluating it at $r_{n}^{*}$, which can be expressed, using~\eq{eq:compr_energy}, as
\be
\label{eq:rate_allocation}
U_{n}(t)r^{*}_{n} + (\theta_{n} - E_{n}(t))\alpha_{n}r^{*}_{n} - r_{n}^{*} \sum_{m\text{: }n \in \X_{m}} \lambda^{*}_{m}  \leq 0.
\ee
From~\eq{eq:energy_bound},~\eq{eq:bound_lambda} and~\eq{eq:rate_allocation}, we further obtain:
\be
U_{n}(t) \leq \sum_{m\text{: }n \in \X_{m}} \lambda^{*}_{m} \leq \gamma_{n} V ,
\ee
which implies that a node $n$ receives exogenous data from outside the network only when $U_{n}(t) \leq \gamma_{n} V$. Hence, recalling that $R_{n}(t) \leq \Rmax$, we obtain the desired result $U_{n}(t+1) \leq \gamma_{n} V + \Rmax$.

Finally, if a node $n$ receives both endogenous and exogenous data, we have from~\eq{eq:cond_endo} that $U_{n}(t) \leq \gamma_{n} V  -\lmax \mumax$. Since a node $n$ can receive at most $\lmax \mumax$ bits per channel use of endogenous data and $\Rmax$ bits per channel use of exogenous data, we have the desired inequality $U_{n}(t+1) \leq \gamma_{n} V + \Rmax$, which completes the proof of part 1).

\noindent 2) To prove the claim, we need to show that if 
\be
\label{eq:minimum_energy}
E_{n}(t) < \alpha_{n}\Rmax + \Pmax, 
\ee
then the following two conditions must be satisfied:
\ben[a)]
\e the Rate-Distortion problem~\eq{eq:rate_dist_alloc} is minimized by choosing $R_{n}(t) = r_{n}^{*} = 0$ (which implies $\Pcn(t)=0$) for all $n \in \N$; 
\e the Power Allocation problem~\eq{eq:power_alloc} selects a power matrix $\P(t)$ such that $P_{n}(t) =0$ for all $n \in \N$.
\een
From Lemma~\ref{lem:bound_lambda}, and in particular from~\eq{eq:rn_star}, condition a) is verified if 
\be
\begin{split}
\label{eq:compres_energy_minim}
U_{n}(t) r_{n}& - (E_{n} - \theta_{n}) P^{c}_{n}(r_{n}) \\
&- r_{n} \sum_{m\text{: }n \in \X_{m}} \lambda_{m}^{*} > 0,\ \text{for all } r_{n} > 0,
\end{split}
\ee
where we recall that ${\bm \lambda}^{*}$ is any optimal dual vector of problem~\eq{eq:dual_ratedist2}. This is proved by the following inequalities:
\bea
&& \hspace{-2.5cm} U_{n}(t) r_{n} - (E_{n} - \theta_{n}) \alpha_{n}r_{n} - r_{n} \sum_{m\text{: }n \in \X_{m}} \lambda_{m}^{*} \nonumber \\ 
&& \hspace{-2.5cm} \;\;\;\; > U_{n}(t) r_{n} + \frac{\gamma_{n}}{\beta_{n}}V\alpha_{n}r_{n} - r_{n} \sum_{m\text{: } n \in \X_{m}} \lambda_{m}^{*} \nonumber \\
&& \hspace{-2.5cm} \;\;\;\; \geq U_{n}(t) r_{n} + \frac{\gamma_{n}}{\beta_{n}}V \alpha_{n}r_{n} - r_{n} \gamma_{n}V \nonumber \\
&& \hspace{-2.5cm} \;\;\;\; = U_{n}(t) r_{n} + \gamma_{n}V \frac{(\alpha_{n}- \beta_{n}) r_{n}}{\beta_{n}} \geq 0 \nonumber
\eea
where the first inequality follows from~\eq{eq:minimum_energy} and the assumption of Theorem~\ref{thm:performance} that $\theta_{n} = \frac {\gamma_{n}}{\beta_{n}} V + \alpha_{n}\Rmax + \Pmax$; the second from~\eq{eq:bound_lambda}; and the last inequality follows from $U_{n}(t) \geq 0$, $r_{n} > 0$ and from the definition of $\beta_{n}$. This proves~\eq{eq:compres_energy_minim} and thus that condition a) is satisfied if~\eq{eq:minimum_energy} holds.   

\indent To prove b) we first note that the bound~\eq{eq:data_bound} implies that the weight~\eq{eq:link_weight} satisfies the inequality
\bea
\label{eq:bound_W}
W_{n,m}(t) &=& \max \{ U_{n}(t) - U_{m}(t) - \delta, 0\}	\nonumber \\ 
&\leq& \gamma_{n} V -\lmax \mumax ,
\eea
for all $(n,m) \in \L$ and for all time $t$. 
We now show by contradiction that condition b) holds when~\eq{eq:minimum_energy} is satisfied. To this end, assume that the power allocation vector $\P^{*}$ that maximizes~\eq{eq:power_alloc} at time $t$ is such that some entry $P^{*}_{n,m}$ is positive. Starting from $\P^{*}$, we now obtain a new power allocation vector $\P$, in which we set $P_{n,m} = 0$. Clearly, the power matrix $\P$ is also feasible. We demonstrate that the objective function of~\eq{eq:power_alloc} when evaluated at $\P^{*}$ is smaller than at $\P$, thus leading to a contradiction. Denoting as $G(\P)$ the objective function of~\eq{eq:power_alloc}, this is shown by the following inequalities: 
\bea
&& \hspace{-0.7cm} G(\P^{*}) - G(\P) = \nonumber \\ 
&& \hspace{-0.7cm} \;\; =\sum_{n \in \N} \sum_{l \in \N \setminus n} [\Cap_{n,l}(\P^{*},\S(t))- \Cap_{n,l}(\P,\S(t))] W_{n,l}(t) \nonumber \\
&& \hspace{-0.7cm} \;\;\;\;\; + (E_{n}(t)-\theta_{n})P^{*}_{n,m} \nonumber \\
&& \hspace{-0.7cm} \;\; \leq \Cap_{n,m}(\P^{*},\S(t)) W_{n,m}(t) + (E_{n}(t)-\theta_{n})P^{*}_{n,m} \nonumber \\ 
&& \hspace{-0.7cm} \;\; \leq \Cap_{n,m}(\P^{*},\S(t)) (\gamma_{n} V -\lmax \mumax) + (E_{n}(t)-\theta_{n})P^{*}_{n,m} \nonumber \\ 
&& \hspace{-0.7cm} \;\; \leq (\gamma_{n} V -\lmax \mumax)\xi P^{*}_{n,m} + (E_{n}(t)-\theta_{n})P^{*}_{n,m}  \nonumber \\
&& \hspace{-0.7cm} \;\; < (\gamma_{n} V -\lmax \mumax)\xi P^{*}_{n,m} - \frac{\gamma_{n}}{\beta_{n}}V P^{*}_{n,m} < 0 , \nonumber
\eea
where the first inequality derives from $\mu_{n,l}(\P^{*},\S(t)) - \mu_{n,l}(\P,\S(t)) \leq 0$ for all $l \neq m$ ({\it Property 2}), the second from~\eq{eq:bound_W}, the third from {\it Property 1} and the fourth from~\eq{eq:minimum_energy}.
This shows that $\P^{*}$ is not optimal for~\eq{eq:power_alloc}, thus leading to a contradiction, which completes the proof of 2).

\noindent 3) The proof of 3) is a relatively simple application of the general theory of~\cite{Georgiadis06}\cite{Neely10}. The details are provided in the following for completeness. We first define the standard one-slot conditional Lyapunov Drift-plus penalty of the queues $\E(t)$ and $\U(t)$. To this end, we define $Z_{n}(t) = \left ( U_{n}(t), E_{n}(t) - \theta_{n} \right )$ and the corresponding vector $\Z(t) = \left ( \U(t), \E(t) -\thetabf \right )$. Following the standard definition~\cite{Neely10}, the quadratic perturbed Lyapunov function is given by
\bea
L(\Z(t)) &=& \frac{1}{2} \sum_{n=1}^{N} || Z_{n}(t) || ^{2} \nonumber \\ 
&=& \frac{1}{2} \sum_{n=1}^{N} (U_{n}(t))^{2} + \frac{1}{2} \sum_{n=1}^{N}(E_{n}(t) -\theta_{n})^{2} \nonumber \\
&=& L(\U(t)) + L(\E(t)-\thetabf),
\eea
while the one-slot conditional Lyapunov drift $\Delta(\Z(t))$ is
\be
\Delta(\Z(t)) = \Exp \left [ L(\Z(t+1))-L(\Z(t)) | \Z(t) \right ].
\ee
The proof of the following lemma can be found in Appendix~\ref{sec:proof_lem:drift}.
\lem{\label{lem:drift} Under any feasible policy for problem~\eq{eq:opt_probl} we have the inequality
\begin{eqnarray}
\label{eq:lem_drift}
&& \hspace{-0.5cm} \Delta(\Z(t)) \leq  \widetilde{B} + \sum_{n \in \N} U_{n}(t)\Exp [-\muout(t) + \muin(t) \nonumber \\
&& \hspace{-0.5cm} \;\;\;\;\;\; + R_{n}(t) \vert \Z(t) ]  + \sum_{n \in \N} (E_{n}(t) - \theta_{n}) \Exp \Big [- \Pcn(R_{n}(t)) \nonumber \\
&& \hspace{-0.5cm} \;\;\;\;\;\; - P_{n}(t) + \Htilde(t) \vert \Z(t)\Big ],
\end{eqnarray}
with $\widetilde{B} = N \left ( \mumax(\mumax + \Rmax) + \RmaxSq/2 \right )+ N/2 ( \HmaxSq +\PcmaxSq + \PmaxSq+2\Pcmax\Pmax )$.
}

\begin{figure*}
\begin{eqnarray}
\label{eq:drift+penalty}
&& \Delta(\Z(t)) +V\Exp \left [ \sum_{n \in \N}\fn(D_{n}(t))\Big \vert \Z(t) \right ] \leq \widetilde{B} +\sum_{n \in \N}(E_{n}(t) - \theta_{n})\Exp \left [\Htilde_{n}(t) \vert \Z(t)\right ] \nonumber \\
&& \;\;\;\;\;\; + \Exp \Bigg [ \sum_{n \in \N}( U_{n}(t) R_{n}(t) - (E_{n}(t)-\theta_{n})\Pcn(R_{n}(t)) + V \fn(D_{n}(t))) \Big \vert \Z(t)\Bigg ] \nonumber \\
&& \;\;\;\;\;\; - \Exp \Bigg [ \sum_{n \in \N} \Bigg ( \sum_{m\text{: } (n,m) \in \L} \Cap_{n,m}(\P(t),\S(t)) ( U_{n}(t) - U_{m}(t)) +  (E_{n}(t)-\theta_{n})P_{n}(t) \Bigg ) \Bigg \vert \Z(t) \Bigg ] . 
\end{eqnarray}
\end{figure*}

\indent The proposed policy is based on the minimization of the drift-plus-penalty function~\cite{Georgiadis06}\cite{Neely10} $\Delta(\Z(t)) + V\Exp \left [ \sum_{n \in \N}\fn(D_{n}(t)) \Big \vert \Z(t) \right ]$. This amounts to finding a policy that minimizes the right-hand side of \eq{eq:drift+penalty} (where the inequality follows from~\eq{eq:lem_drift}). Minimization of~\eq{eq:drift+penalty} is done with respect to $(\R(t),\D(t),\Htildebf(t),\P(t))$ for the given $\left ( \S(t), \O(t), \H(t), \U(t), \E(t)\right )$ under the constraints~	\eq{eq:rates_dist} and $0 \leq R_{n} \leq \Rmax$, $\Dmin \leq D_{n} \leq \Dmax$, as per definition of policy in Section~\ref{sec:optimization_problem}. It is now not difficult to see that, similar to~\cite{Huang10}, by Lagrangian relaxation of the constraints~\eq{eq:rates_dist}, the dual function of the said minimization problem, when considering fixed $\O(t) = \o_{i}$, $\S(t)= \s_{j}$, $\H(t) = \h_{k}$ and fixed queue lengths $(\U(t),\E(t))$, is given by $\tilde{\dual}({\bm \lambda}^{(\o_{i})}) = \dual_{\o_{i},\s_{j},\h_{k}}({\bm \lambda}^{(\o_{i})},\U(t),\E(t)-\thetabf)$ as defined in~\eq{eq:dual_function_oneState}. Note that the Lagrange multipliers ${\bm \lambda}^{(\o_{i})}$ are associated to the constraints~\eq{eq:rates_dist}. Moreover, by convexity and Slater's conditions, we have that strong duality holds, and thus the minimum of~\eq{eq:drift+penalty} equals $\tilde{\dual}({\bm \lambda}^{(\o_{i})})$ for a given value ${\bm \lambda}^{(\o_{i})} = {\bm \lambda}^{(\o_{i})*}$.

\indent From the discussion above, the minimum of the right-hand side of the bound~\eq{eq:drift+penalty} equals
\bea
&& \hspace{-2cm} \widetilde{B} + \Exp[\dual_{\o_{i},\s_{j},\h_{k}}({\bm \lambda}^{(\o_{i})*},\U(t) , \E(t)-\thetabf ) \vert \Z(t)] = \nonumber \\
&& \hspace{-2cm} \;\;\;\; = \widetilde{B} + \dual({\bm \lambda}^{*},\U(t) , \E(t)-\thetabf )
\eea
for some ${\bm \lambda}^{*} \in \mathbb{R}^{L(2^{N}-1)}_{+}$ (${\bm \lambda}^{*}$ collects all ${\bm \lambda}^{(\o_{i})*}$). But by Theorem~\ref{thm:optimal}, we have that
\be
V F_{0}^{*} \geq \dual({\bm \lambda}^{*},\U(t) , \E(t)-\thetabf ).
\ee
From~\eq{eq:drift+penalty}, we can now write that for the considered policy that minimize~\eq{eq:drift+penalty}, we have the inequality
\be
\Delta(\Z(t)) + V\Exp \left [ \sum_{n \in \N}\fn(D_{n}(t))\Bigg \vert \Z(t) \right ] \leq \widetilde{B} + V F_{0}^{*}.
\ee
Moreover, taking expectation over $\Z(t)$ and summing the above over $t=0,\ldots,T-1$, we have:
\bea
&& \hspace{-1cm} \Exp[L(\Z(T))-L(\Z(0))] + V \sum_{t=0}^{T-1} \Exp \left [ \sum_{n \in \N}\fn(D_{n}(t)) \right ] \nonumber \\
&& \hspace{-1cm} \;\;\;\; \leq T \widetilde{B} + T V F_{0}^{*}.
\eea
Rearranging the terms, using the fact that $L(\Z(t)) \geq 0$ and $L(\Z(0))=0$, dividing both sides by $VT$, and taking the limsup as $T \to \infty$, we get:
\be
\label{eq:final_bound}
\limsup \sum_{n\in \N} \frac{1}{T} \sum_{t=0}^{T-1} \Exp [ \fn(D_{n}(t))] \leq F_{0}^{*} + \frac{\widetilde{B} }{V}.
\ee
This shows that our policy satisfies the desired claim.

\indent It remains to be discussed whether the proposed policy does indeed minimize~\eq{eq:drift+penalty}. It can be seen, similar to~\cite{Huang10} that the proposed policy minimizes a modified version of~\eq{eq:drift+penalty} where $(U_{n}(t)-U_{m}(t))$ is replaced by $\max\{ U_{n}(t)-U_{m}(t) - \delta , 0 \}$ (cf.~\eq{eq:link_weight}). Moreover, when $(\theta_{n} - E_{n}(t)) < H_{n}(t)$, we harvest a reduced amount of energy. This implies that the right-hand side of~\eq{eq:drift+penalty} under the proposed policy is generally larger than with the policy discussed above that minimizes the right-hand side of~\eq{eq:drift+penalty}. However, the loss is at most
\be
0 \leq \sum_{n \in \N} \sum_{m\text{: }(n,m) \in \L} \mu_{n,m}(t) \delta \leq N \delta \lmax \mumax
\ee
for the power allocation part of the algorithm, and 
\be
0 \leq \sum_{n \in \N} (\theta_{n} - E_{n}(t))(H_{n}(t) - (\theta_{n} - E_{n}(t))) \leq \frac{N \HmaxSq}{4}
\ee
for the energy harvesting. This shows that~\eq{eq:final_bound} also holds for the proposed policy as long as we substitute $\widetilde{B}$ with $B$. This concludes the proof.
\end{IEEEproof}

\section{Proof of Lemma~\ref{lem:bound_lambda}}
\label{sec:proof_lem:bound_lambda}
\begin{IEEEproof}
Let ${\bm \lambda}^{*}$ and ${\bm \upsilon}^{*}$ be an optimal solution of the dual problem~\eq{eq:dual_ratedist2}, and $\r^{*} = [r_{1}^{*}, \ldots, r_{N}^{*}]$ and $\d^{*} = [d_{1}^{*}, \ldots, d_{N}^{*}]$ be an optimal solution of the (primal) problem~\eq{eq:rate_dist_alloc}. Existence of $(\r^{*},\d^{*})$ and $({\bm \lambda}^{*}, {\bm \upsilon}^{*})$ is guaranteed by Weierstrass theorem~\cite[Proposition 2.1.1]{bertsekasConvexAnalysis} and by Slater's condition~\cite[Proposition 3.5.4, part a)]{bertsekasConvexAnalysis}. By~\cite[Proposition 6.1.1]{bertsekasConvexAnalysis}, the following conditions must be satisfied by $\d^{*}$ and $({\bm \lambda}^{*}, {\bm \upsilon}^{*})$: primal feasibility, namely $d_{n}^{*} \leq \Dmax$, and the complementary slackness conditions $\upsilon_{n}^{*}(d_{n}^{*}-\Dmax)=0$ for all $n \in \N$, and $(\r^{*},\d^{*}) = {\operatorname{argmin\ }} \Lag(\r,\d,{\bm \lambda}^{*},{\bm \upsilon}^{*})$ where the minimization is taken under the constraints $d_{n} \geq \Dmin$ and $0 \leq r_{n}^{*} \leq \Rmax$ for all $n \in \N$. From~\eq{eq:lagr_ratedist2}, the given conditions imply that
\be
\begin{split}
V\fn&(\Dmax) - \log(\Dmax) \sum_{m\text{: }n \in \X_{m}} \lambda_{m}^{*}\\ 
&- \left ( V \fn(d^{*}_{n}) - \log(d^{*}_{n}) \sum_{m\text{: }n \in \X_{m}} \lambda_{m}^{*} \right ) \geq 0 ,
\end{split}
\ee
must be satisfied. This is because the Lagrangian $\Lag(\r,\d,{\bm \lambda}^{*},{\bm \upsilon}^{*})$ when evaluated at $d_{n}=d_{n}^{*}$ should be no larger than for $d_{n} = \Dmax$. We thus have the inequalities
\bea
&& \hspace{-1cm} \sum_{m\text{: } n \in \X_{m}} \lambda_{m}^{*} \leq \frac{\fn(d^{*}_{n})-\fn(\Dmax)} {\log(d^{*}_{n}/\Dmax)} V \nonumber \\
&& \hspace{-1cm} \;\;\;\;\;\;\;\; \leq \sup_{\Dmin \leq d_{n} \leq \Dmax} \left [ \frac{\fn(d_{n})-\fn(\Dmax)} {\log(d_{n}/\Dmax)} \right ] V = \gamma_{n} V, \nonumber 
\eea
where the second inequalities follows since $\Dmin \leq d_{n}^{*} \leq \Dmax$ and the third from the definition of $\gamma_{n}$.
\end{IEEEproof}

\section{Proof of Lemma~\ref{lem:drift}}
\label{sec:proof_lem:drift}
\begin{IEEEproof}
First, let us consider the time evolution of the data queue $U_{n}(t)$ of a generic node $n$. By squaring both sides of~\eq{eq:data_queue} and using the fact that for any $x \in \mathbb{R}$, $(\max(x,0))^{2} \leq x^{2}$, we have:
\begin{eqnarray}
&& \hspace{-1cm} ( U_{n}(t+1))^{2} - ( U_{n}(t) )^{2} = ( \max ( U_{n}(t) - \muout(t), 0 ) \nonumber \\
&& \hspace{-1cm} \;\;\;\;\;\;\;\; + \muin(t) + R_{n}(t) )^{2} - ( U_{n}(t) )^{2} \nonumber \\
&& \hspace{-1cm} \;\;\;\; \leq (\muout(t))^{2} + (\muin(t) + R_{n}(t))^{2} - 2\muout(t) (\muin(t) \nonumber \\
&& \hspace{-1cm} \;\;\;\;\;\;\;\; + R_{n}(t)) + 2U_{n}(t)(- \muout(t) + \muin(t) + R_{n}(t)) \nonumber \\
&& \hspace{-1cm} \;\;\;\; \leqÊ(\muout(t))^{2} + (\muin(t) + R_{n}(t))^{2} \nonumber \\
&& \hspace{-1cm} \;\;\;\;\;\;\;\; + 2U_{n}(t)(- \muout(t) + \muin(t) + R_{n}(t)).
\end{eqnarray}
By defining $B_{U} = \mumax(\mumax + \Rmax) + \RmaxSq/2$, we then see that:
\bea
&& \hspace{-2cm} \frac{1}{2}[ (U_{n}(t+1))^{2} + (U_{n}(t))^{2} ] \nonumber \\
&& \hspace{-2cm} \;\;\;\;\;\; \leq B_{U} + U_{n}(t)[- \muout(t) + \muin(t) + R_{n}(t)].
\label{eq:lyap_data}
\eea
Similarly, let us consider the perturbed evolution of the energy queue $E_{n}(t)$. By squaring both sides of~\eq{eq:energy_queue} we have:
\begin{eqnarray}
&& \hspace{-1.5cm} ( E_{n}(t+1) - \theta_{n})^{2} - ( E_{n}(t) - \theta_{n})^{2} = \nonumber \\
&& \hspace{-1.5cm} \;\;\;\; =( E_{n}(t) - \Pcn(R_{n}(t)) - P_n(t)+ \Htilde_{n}(t) - \theta_{n} )^{2}  \nonumber \\ 
&& \hspace{-1.5cm} \;\;\;\;\;\;\;\; - \left ( E_{n}(t) - \theta_{n} \right )^{2}  \nonumber \\
&& \hspace{-1.5cm} \;\;\;\; = (- \Pcn(R_{n}(t)) - P_n(t)+ \Htilde_{n}(t) )^{2}  \nonumber \\
&& \hspace{-1.5cm} \;\;\;\;\;\;\;\; + 2(E_{n}(t)- \theta_{n})(- \Pcn(R_{n}(t)) - P_n(t)+ \Htilde_{n}(t)).
\end{eqnarray}
By defining $B_{E} = \frac{1}{2}( \HmaxSq +\PcmaxSq + \PmaxSq+2\Pcmax\Pmax )$, we then see that:
\begin{eqnarray}
\label{eq:lyap_energy}
&& \hspace{-1.5cm} \frac{1}{2}\left [ (E_{n}(t+1) - \theta_{n})^{2} -  (E_{n}(t) - \theta_{n})^{2} \right ] \nonumber \\
&& \hspace{-1.5cm} \leq B_{E} + (E_{n}(t)- \theta_{n})(- \Pcn(R_{n}(t)) - \Pout(t) + \Htilde_{n}(t)).
\end{eqnarray}
Now by summing~\eq{eq:lyap_data} and~\eq{eq:lyap_energy} over all $n \in \N$, and by defining $\widetilde{B} = N(B_{U} + B_{E}) = N \left ( \mumax(\mumax + \Rmax) + \RmaxSq/2 \right )+ N/2 ( \HmaxSq +\PcmaxSq + \PmaxSq+2\Pcmax\Pmax )$, we have:
\bea
&&\hspace{-1.25cm} L(\Z(t+1))-L(\Z(t)) \leq \widetilde{B} \nonumber \\ 
&&\hspace{-1.25cm} \;\;\;\;+\sum_{n=1}^{N} U_{n}(t)(- \muout(t) + \muin(t) + R_{n}(t)) \nonumber \\
&&\hspace{-1.25cm} \;\;\;\;+ \sum_{n=1}^{N} (E_{n}(t) - \theta_{n})(- \Pcn(R_{n}(t)) - \Pout(t) + \Htilde_{n}(t) ]. 
\eea
Taking the expectation on both sides over the random observation, channel and energy harvesting and conditioning on $\Z(t)$, the lemma follows.
\end{IEEEproof}

\section{Source model}
\label{sec:source_model}
Here we present a simple source model for which we determine numerical results in Section~\ref{sec:results}. Let the source signals measured at sensors in $\N$ be spatially correlated with parameter $\omega$. Since the measurements are Gaussian we can write for the $i$th sensor $X_{i} = \sqrt{\omega}A+\sqrt{1-\omega}B_{i}$, with $A$ and $B_{i}$ independent Gaussian random variables with zero mean and unitary variance. Moreover, we assume that the sink is able to measure $A$ with an accuracy that depends on the rate $R_{d}$ used for acquisition. From standard rate-distortion theory, we have the relationship $R_{d} = I(A;Y)$, where $Y$ is the side information available at the sink. By choosing the optimal test channel $Y=\sqrt{\omega_{d}}A+\sqrt{1-\omega_{d}}C$, where $\omega_{d}$ denotes the correlation between the measurement $Y$ at the sink and $A$ (see, e.g.,~\cite{ElGamal12}), we obtain the equations reported in the text. 

\end{document}